\documentclass[11pt]{article}
\usepackage{fullpage}
\usepackage[english]{babel}
\usepackage[utf8]{inputenc}

\usepackage{fancyhdr}
\usepackage[margin=1in]{geometry}
\usepackage{natbib}
\usepackage{setspace}
\usepackage{graphicx}
\graphicspath{ {./pics/} }

\usepackage{placeins}
\usepackage{float}
\usepackage[toc,page]{appendix}

\usepackage{bm}
\usepackage{amsmath,amssymb,amsthm}
\usepackage{mathtools}
\usepackage{nccmath}

\usepackage{comment}
\newcommand*\samethanks[1][\value{footnote}]{\footnotemark[#1]}

\usepackage{array,multirow,graphicx}
\usepackage[table,dvipsnames]{xcolor}
\usepackage{siunitx,booktabs,caption}
\usepackage[flushleft]{threeparttable}

\usepackage{colortbl} 

\usepackage{tabularray}

\usepackage{makecell}

\usepackage{amssymb}

\usepackage{rotating}

\usepackage[normalem]{ulem}

\usepackage{xurl,hyperref}
\usepackage{url}

\title{
	\vspace{0.5 cm}
	\textbf{Vulnerabilities and capabilities in the EU Automotive industry: Leveraging Input-Output Analysis and Economic Complexity}
}

\vspace{0.4 cm}

\bigskip

\author{
	Lorenzo Cresti\thanks{Enrico Fermi Research Center (CREF), Via Panisperna, 89a, 00184 Roma (RM).
		Corresponding author: \texttt{angelica.sbardella@cref.it}.} \and
	Dario Mazzilli\samethanks \and
	Aurelio Patelli\samethanks \and  
	Angelica Sbardella\samethanks \and
	Andrea Tacchella\samethanks
}

\date{December 2024}

\begin{document}
	
	\maketitle
	\thispagestyle{fancy}
	
	\begin{abstract}
		
		This paper investigates the structural vulnerabilities and competitive dynamics of the EU27 automotive sector, with a focus on the complexity and the fragmentation of production processes across global value chains. Employing a mixed-methods approach, our analysis integrates input-output tables to quantify the sector's reliance on non-EU economic branches, alongside an economic complexity framework to assess the underlying productive capabilities of European countries in automotive-related industries. The findings indicate an increasing dependency on extra-EU suppliers, particularly China, for critical components such as lithium-ion batteries, which heightens supply chain risks. Currently, Eastern European countries—most notably Poland, Czechia, and Hungary—have enhanced their competitiveness in the production of automotive components, surpassing traditional leaders such as Germany. The paper advances the literature by providing a novel, granular list of 6-digit products within the automotive supply chain and offers new insights into the challenges posed by the ongoing electric mobility transition in the European Union, particularly in relation to electric accumulators.
	\end{abstract}
	\bigskip 
	\noindent\textbf{Keywords}: European automotive industry, electric vehicles, value chain vulnerability, input-output, global value chains, economic fitness and complexity, relatedness\\
	\bigskip
	\textbf{JEL classification codes}: F16, F6, J24, L6, O14

	\newpage
	\clearpage
	\begin{spacing}{1.5}
		\section{Introduction}
		
		
		The automotive sector has long been a focal point of social science research due to its profound transformations over time and its crucial role in generating value added and employment. This is especially true for the EU, that has traditionally been an automotive leader and where today the automotive industry share of the whole manufacturing sector real value added is nearly 10\%, providing direct and indirect employment to 13.8 million Europeans. This number represents 8.5\% of EU manufacturing employment and 6.1\% of total EU employment. Additionally, the automotive industry accounts for 4\% of extra-euro area exports.\footnote{See, for instance, the European Central Bank Economic Bulletin Issue 4, 2024 (\url{https://www.ecb.europa.eu/press/economic-bulletin/html/eb202404.en.html}). See also the annual data highlighted in ACEA's automobile industry pocket guide (\url{https://www.acea.auto/publication/the-automobile-industry-pocket-guide-2023-2024/}).}
		
		The transition to electric mobility poses significant challenges for car manufacturers, necessitating a reorganization of production processes and shifts in supply chain inputs, technologies, and productive capabilities. The most notable changes center on engines and related components, with a growing demand for batteries, electrochemical cells, and critical raw materials.\footnote{Battery materials such as lithium, cobalt, magnesium, nickel, and graphite are concentrated in a few countries, heightening supply chain vulnerabilities for car manufacturers \citep{richert2023selected}.}
		These vulnerabilities are compounded by China's dominant position, not only as a key processor of battery materials but also as a leading manufacturer of electric vehicle components, including batteries and related technologies. This dual specialization raises concerns about reliance on Chinese suppliers across multiple stages of the value chain, further amplifying risks for EU manufacturers \citep{altenburg2022china}.
		
		These trends are increasingly straining the EU automotive industry, prompting policymakers to propose provisional tariffs on electric vehicles imported from China.\footnote{\url{https://ec.europa.eu/commission/presscorner/detail/en/statement_24_5041}} This initiative aligns with the broader framework of Open Strategic Autonomy, developed after the COVID-19 pandemic in response to supply chain disruptions and shortages. The framework aims at achieving a new balance between security and competitiveness -- preserving trade openness while mitigating vulnerabilities from economic interdependence. A "de-risking" approach that seeks to mitigate vulnerabilities and supply chain interdependencies while maintaining an open global economy \citep{van2021towards, gehrke2022eu, schmitz2023open}. 
		
		Against the backdrop of supply chain contractions and the electric mobility transition, this paper employs input-output analysis, Economic Complexity, and machine learning to investigate the EU27 automotive industry’s input dependencies, competitiveness, and vulnerabilities.
		To fully capture the sector's transformations, the study moves beyond conventional sectoral perspectives by recognizing that the EU automotive industry's productive capabilities are shaped by the global fragmentation of production along value chains. Manufacturing a car involves assembling thousands of components sourced worldwide, exposing the industry to strategic dependencies and supply chain vulnerabilities. A case in point is electric vehicle batteries, a critical component predominantly sourced through international supply chains, with a significant share currently imported from China. We thus propose a novel empirical strategy that simultaneously analyses supply chain dependencies and export capabilities  integrating granular product-level trade data with inter-sectoral linkages—two dimensions rarely examined together in supply chain studies. To this aim, we propose a novel a microdata-validated list of 63 harmonized system (HS) intermediate products related to motor vehicle production, provided in the Appendix.
		
		In our empirical analysis, first, we examine the industry's dependence on key trade partners and products, highlighting a shift in suppliers through input-output analysis and product-level trade.  China, along with Turkey, South Korea, and Mexico, has gained importance, while countries like the UK and Japan have seen a decline in relevance. Notably, imports of electrical parts, especially electric accumulators, have surged, with China as the EU’s primary supplier.
		Second, using the Economic Fitness and Complexity metric, we evaluate the competitiveness of EU27 countries in automotive-related exports. Our analysis  shows that Eastern European nations—particularly Poland, Hungary, and Czechia—have improved their competitiveness in automotive components, surpassing Germany. Additionally, these countries, along with Italy, display the highest probability of developing future comparative advantages in automotive components.
		Third, we focus on the electric mobility transition: while EU27 exports of electric and hybrid vehicles have grown, internal combustion engine vehicles still dominate, and the region lags behind in battery production capabilities, remaining heavily reliant on lithium-ion battery imports from China. Some EU countries are positioned to become competitive in certain types of batteries, though very few are positioned to become competitive in lithium-ion battery exporters by 2027.
		Finally, we conduct a vulnerability assessment of the EU27 automotive supply chain, combining product-level data on net exposure to extra-EU27 inputs and supply chain import concentration. Products in the Electrical and Electric Parts and Engines and Parts categories are identified as the most vulnerable, with lithium-ion batteries and electric accumulators posing significant risks due to high external dependency and supply concentration. 
		
		The remainder of the paper is structured as follows. In Section \ref{sec:lit} we provide a brief literature review documenting the current features of the European automotive industry. Section \ref{sec:methods} introduces the methods and the data used. Section \ref{sec:results} presents our empirical analysis, and Section \ref{sec:conclusion} concludes.
		
		\section{The European automotive industry: challenges and transformations}\label{sec:lit}
		The automotive industry has long been a strategic sector for the EU’s industrial, economic, and social development \citep{klebaner2022european}. It consistently accounted for a large share of the European manufacturing output, with an important employment absorption capacity within and outside the sector. Historically, it has been a primary driver of growth for the European economy \citep{boranova2022cars,ECcars20,ECcars21,maiza2018analysis}.
		Numerous scholars have explored transformations within the EU's automotive industry, focusing on its recent developments driven by technical change, as well as its changing mode and geography of production. \cite{lung2004changing} and \cite{domanski2009changing}  discuss early changes in the mode of production, while more recent studies focus on evolving industrial strategies. \cite{gaddi2021automotive}, \cite{simonazzi2020future} and \cite{simonazzi2022world} delve into the future and current challenges of the European automotive sector, emphasizing the impact of technological shifts and restructuring dynamics. Other studies have examined the cost dynamics and regional shifts in production, pointing out how different regions have adapted to these transformations \citep{garcia2020analysing, grodzicki2020cost}.
		
		In particular, since the mid-1980s, the industry faced a profound production reorganization due to outsourcing to other branches and offshoring to emerging economies. The work of \cite{pavlinek2017dependent, pavlinek2020restructuring} and \cite{fana2021automotive} illustrates these dynamics, focusing on the role of dependent market economies and the growing importance of Eastern Europe in the automotive supply chain. As global trade increasingly focused on intermediate goods, car producers intensified coordination and collaboration with various supplier firms, leading to increasing vertical integration \citep{russo2023regionalisation, boranova2022cars}. The growth of supply chain specialization in the car industry has largely stemmed from labour cost competitiveness -- i.e.,offshoring with the aim of reducing costs -- and changes in the economic landscape, especially the EU enlargements and the growing importance of Asian industrial economies that enabled multinationals to relocate productive units in Eastern Europe and Asia \citep{frigant2014automotive, frigant2017regionalisation,garcia2020analysing}. 
		While these delocalisation strategies have yielded some efficiency gains and costs reductions \citep{boranova2022cars}, they have also heightened vulnerability to supply chain disruption shocks and altered countries' employment absorption capacities. This critical aspect of global value chains has become particularly pronounced amid  recent crisis stemming from the global pandemics, geopolitical tensions and the urgency of the green transition. 
		Vulnerabilities are especially acute for industries reliant on highly specialised intermediate inputs supplied by a limited number of providers along value chains. \\
		Fragmentation of production  has also occurred within Europe borders. The European automotive, like many other industries, exhibits a core-periphery structure \citep{celi2017crisis,gerocs2019relocation,pavlinek2022relative}, characterised by a central-continental lead area, with Germany at its core, and various peripherical regions specialising as suppliers. For instance, Eastern EU countries have historically provided labour intensive components to Germany but are now increasing the technological sophistication of their goods, indicating potential upgrades within the automotive value chain \citep{gaddi2021automotive,gaspar2023changes}. Additionally, EU neighbouring countries like Morocco and Turkey have emerged as new suppliers of automotive components for European manufacturers, also showing signs of upgrading\citep{haddach2017moroccan,mordue2022upgrading}.
		
		Despite sourcing over 60\% of intermediate inputs within Europe, the automotive industry relies more heavily on foreign inputs than other sectors \citep{boranova2022cars}, making it particularly susceptible to supply chains disruption.  \cite{gracia2017network} document an increasing dependence on extra-regional markets for both motor vehicle/car foreign demand and imports of components. Currently, the main challenge facing the EU's automotive sector is the rise of specialised extra-EU suppliers for strategic products related to electric mobility. 
		This shift is prompting a reconfiguration of the sector as Germany strives to maintain its central position while countries like France and Italy experience relative decline \citep{gracia2024analysis}.\footnote{For insights into how the green and digital transitions are reshaping the broader automotive ecosystem see \cite{dechezlepretre2023green}.}
		
		China has emerged as not only the largest car producer globally but also as the primary producer and supplier of essential components for electric vehicles (EVs), particularly batteries.  This dependency poses significant risks since no other country can easily replace China as a supplier.\footnote{In-depth studies of China's catch-up and success in the automotive industry can be found in \cite{altenburg2022china} and \cite{yeung2019made}, while further  insights into the implications for European automotive in \cite{maiza2018analysis}.} \\
		
		As any other massive value chain reorganization, the transition to EVs brings structural changes that may profoundly impact employment absorption capacity and the capabilities embedded in the workforce of dismantled automotive factories \citep{cetrulo2023automation,cresti2023weak,cresti2023italy,pavlinek2020restructuring}. According to \cite{cetrulo2023automation}, while losses in production and employment within the European auto industry are expected to be partially offset by battery manufacturing, some recovery may occur through the electrification of the sector. 
		However, this will depend on EU car producers' ability to reorient productive capabilities, secure critical raw materials (CRMs) and battery components, or develop alternative battery technologies, while competing with other non-EU countries that have entered earlier in the race and already built substantial expertise for manufacturing EV components \citep{gracia2024analysis}.\\
		Overall, there is a clear shift from competitiveness to vulnerability in discussions surrounding the EU's automotive sector and in general the EU's productive structure \citep{bontadini2023european,guarascio2024between}. Analysts highlight not only the EU automotive  declining global position -- signalled by decreasing vehicle production alongside rising car imports -- but also growing concerns about value chain organization. The EU's lack of control over the upstream stages such as CRM mining and processing and battery manufacturing exacerbates these vulnerabilities.
		
		These observations underscore the need for in-depth empirical research into supply chain vulnerabilities while also considering export competitiveness. Our study contributes to the literature by focusing on extra-EU trade dynamics  -- both import and exports -- and providing evidence of input dependencies along with competitiveness in final and intermediate automotive products.

		\section{Methods and data: EFC and Input-Output Analysis}\label{sec:methods}
		We propose an approach that takes advantage of a bridge between the product-level information contained in the trade data-set UN-COMTRADE and that on sectoral output in the OECD-ICIO tables. 
		Inter-sectoral linkages are captured using Input-Output tables \citep{miller2022input}, which measure the flows of inputs between countries and generally 2-digit sectors. These tables have been used to understand countries’ participation in global value chains or global production networks, emphasising backward or forward linkages, and measuring degrees of upstreamness or downstreamness \citep{antras2013organizing,antras2018measurement,ponte2019handbook}. In our case, this helps addressing the interdependencies between European automotive sectors and other EU and extra-EU economic branches.
		
		Product-level data provides a more granular perspective of trade between countries, beyond coarse-grained sectoral aggregations. This data has been largely used in the Economic Complexity framework (EC) \citep{hausmann2007structure,hidalgo2009building,tacchella2012new}, which has enabled more accurate analysis and forecasting of economic growth compared to conventional methods \citep{cristelli2013measuring,tacchella2018dynamical}, providing scholars and policymakers with a tool to forecast the trajectories of economic diversification at both country and regional levels. 
		EC considers a fine-grained and structural vision of the productive possibilities of the economy, emphasising the importance of export specialisation patterns for long-run growth \citep{hausmann2003economic,hausmann2006structural,hausmann2007structure}. A country’s trade specialisation profile can in fact be regarded as a reflection of its underlying productive capabilities, envisioned as the skills that enable an economy to expand into new production requirements and adopt new technologies \citep{dosi2010technical,cimoli2009industrial}.
		Building on the capability-based-theory of the firm \citep{costa2021organizational,dosi1984technical,teece1994understanding}, in the EC perspective, the notion of capabilities is reframed as macro location-based  attributes\citep{sbardella2018role}: all the elements that affect a country’s ability to operate modern and large-scale businesses, including their political and social characteristics, such as physical and human capital endowments, institutions, skills, technologies, and more  \citep{abramovitz1986catching,hausmann2003economic}. The main difficulty with the empirical analysis of capabilities is that they are unobservable. It is therefore necessary to develop a methodology to obtain information about them from observable variables. \cite{hausmann2007you} devised an approach to infer the capabilities of an economy from a country’s export specialisation profile. The rationale behind their approach is that the production of complex goods requires the existence of a broad set of advanced skills, as well as the ability to combine them effectively. The production of complex goods that are exported through international markets therefore conveys relevant information about the underlying, unobservable capabilities of the economic system. More details on the approach are provided in the following subsections.
		
		The need to bridge these two approaches arises from their respective drawbacks  and potential complementarity: EC assumes that 'what you export' is a proxy of 'what you produce competitively', and thus of the capabilities you are endowed with, neglecting the fragmentation of production and the import of inputs from abroad. For instance, a country may specialise in exporting final commodities like cars while importing the most technologically advanced and capabilities embedded components essential for their production. This scenario can lead to misinterpretation, where a country's success in exporting vehicles is mistakenly seen as proof of its ability to manufacture all components, including those embodying cutting-edge technology and capabilities.  Recent research stresses the importance of accounting for the import dimension of trade when assessing a country's economic capabilities in the EC framework \citep{ferrarini2015product,ferrarini2016production,bam2020input,koch2021yet,hernandez2023functional,karbevska2023mapping,sebestyen2024resolving}. 
		In contrast, Input-Output Analysis provides precise information on how sectors or countries depend on inputs and where their intermediate or final exports go. However, it is limited by its use of broad 2-digit sector level of analysis. This high-level aggregation prevents researchers from exploring finer details of sector interactions or product-specific relationships.  For instance, in the automotive industry, this limitation makes it difficult to analyze the complexity of specific car parts a country imports to produce its exported vehicles, hindering a more nuanced understanding of the technological sophistication embedded in these component-level trade flows.
		
		\subsection{Economic Fitness and Complexity}
		The basic intuition of the Economic Fitness and Complexity (EFC) approach adopted in this report is that specific products are important because they constitute different learning opportunities and therefore development possibilities. EFC explicitly builds on the heterogeneity of the interactions between different economic actors, assuming that the level of technological and scientific knowledge of a geographical area -- be it a country, a region or a city -- cannot be reached at an intensive level: knowledge grows by accumulation, hence by adding new and different elements to existing capabilities.
		
		In a nutshell, EFC is a recursive algorithm \citep{tacchella2012new} based on the observation of an empirical network connecting countries to the products they export with a comparative advantage. The output of this algorithm defines two indices: Fitness (F), a relative metric for the competitiveness of the productive system of countries, and Complexity (Q), a measure of the sophistication of the products they export. The only input of the EFC  algorithm is the specialisation matrix $M_{cp}$, a binary matrix whose elements reflect the exports by countries (in rows) of products (in columns, and whose entries take value 1 if country $c$ is specialised in product $p$, and 0 otherwise. Specialisation is computed with Balassa's Revealed Comparative Advantage (RCA) \citep{Balassa1965}:
		\begin{equation}
			\label{eq:rca_gamma_alpha}
			RCA_{cp}= \dfrac{W_{cp}}{\sum_{c^{'} p} W_{c^{'} p}}
			/
			\dfrac{\sum_{cp^{'}} W_{cp^{'}}}{\sum\limits_{c^{'} p^{'}} W_{c^{'} p^{'}}}.
		\end{equation}
		where $W_{cp}$ is the export volume of country $c$ in product $p$ in USD, and the matrix element of $M_{cp}$ is defined as follows:
		\begin{eqnarray}\label{eq:m_gamma_alpha}
			M_{cp}=
			\begin{cases}
				1 & \text{ if $RCA_{cp} \geq 1$}\\ 
				0 & \text{otherwise.}
			\end{cases}
		\end{eqnarray}
		Based on the observation of the nested structure of the trade country-product network\citep{mariani2019nestedness,mariani2024ranking}, $F_c$ and $Q_p$ are then defined as: 
		\begin{equation}
			\left\{\begin{array}{r@{}l@{\qquad}l}
				\widetilde{F}_c^{(n)}=\sum_p M_{cp} Q_p^{(n-1)} &\qquad & F_c^{(n)}=\frac{\widetilde{F}_c^{(n)}}{\langle\widetilde{F}_c^{(n)}\rangle_g} \\ \\
				\widetilde{Q}_p^{(n)}=\frac{1}{\sum_{c} M_{cp} \frac{1}{F_{c}^{(n)}}}  &\qquad & Q_p^{(n)}=\frac{\widetilde{Q}_p^{(n)}}{\langle\widetilde{Q}_p^{(n)}\rangle_a}
			\end{array}\right.
			\label{eq:fitness}
		\end{equation}
		
		with $F^{(0)}_c = 1$ and $Q^{(0)}_p = 1$.
		
		The traditional Fitness measure is relative, meaning that if one country's Fitness increases, another's must decrease within the same year. This relativity makes it challenging to compare Fitness scores across different years. To address this limitation, we adopt the approach proposed by \cite{mazzilli2024equivalence}, which introduces a 'dummy' country into the calculations. This dummy country is designed to be specialized in every product and is assigned a constant Fitness value of 1. By including this idealized country in our computations, we create a consistent benchmark against which all other countries' Fitness can be measured. This method allows for more meaningful comparisons of Fitness scores across different years, as the dummy country serves as a fixed reference point in all periods.

		The Fitness metric can be calculated at a more granular level, focusing on specific subsets of products rather than traditional industrial sectors. This approach, known as Sector Fitness,\footnote{In the context of Sector Fitness, the term \textit{sector} refers to the definition of sets and subsets, not specifically to industrial sectors.} allows for a targeted analysis of a country's diversification and product uniqueness within particular areas of interest. We propose computing Fitness on new aggregates based on product similarities in terms of production stage (e.g., intermediate or final goods) or supply chain membership, rather than standard economic branch classifications.
		
		The Fitness metric can also be adapted to focus on specific subsets of products, extending beyond traditional industrial sector boundaries \citep{barbieri2023regional,pugliese2021economic}. This approach, known as Sector Fitness\footnote{In the context of Sector Fitness, the term \textit{sector} refers to the definition of a subset of products, not specifically to industrial sectors.}, allows for a more targeted analysis of a country's competitiveness in specific domains of the global market, as we do here for automotive products -- more details on the selection of automotive commodities is provided in Section \ref{ssec:autom_prods} and in the Appendix. 
		To implement this, we first compute the Fitness algorithm across all products, which generates complexity scores for each product. Then, we select a subset of target automotive products and use only their complexity scores in the final Fitness computation. This method enables us to calculate Fitness for new proposed aggregates based not on standard economic branch classifications, but on product similarities in terms of production stage (e.g., intermediate or final goods) or belonging to a specific supply chain.
		
		\subsection{Product-Progression Probability}
		We also leverage the Product Progression approach, which builds on the Economic Complexity conceptual framework to produce product-level forecasts of countries’ competitiveness. This method is rooted in the concept of product space \citep{hidalgo2009building,zaccaria2014taxonomy}, which posits that the similarity between products can be inferred from the capabilities required to achieve competitive advantage in their production. The foundational approach assesses product similarity by observing the patterns of co-occurrence of pairs of products in countries' export baskets, indicating shared capabilities in their production. This allows us to a quantify a country's relatedness to a product based on how many other products requiring similar capabilities it already exports with comparative advantage.

\begin{figure}[h]
			\centering
			\includegraphics[scale=1,trim={0cm 0cm 0cm 0cm}]{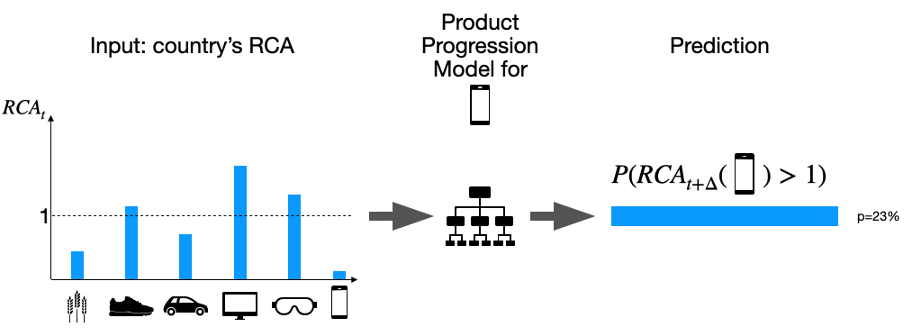}
			\caption{Product Progression network}
			\label{productprogression}
		\end{figure}

		The estimation of relatedness can be refined by considering more complex and non-linear relationships between products, using machine learning techniques to enhance prediction accuracy \citep{tacchella2023relatedness}. This methodology is called Product Progression and it is schematically illustrated in Figure \ref{productprogression}.
		Specifically, we train Tree-Based models (approximately 5000, one for each product) to predict the probability of a country exporting a given product with $\text{RCA}>1$ after a 5-year period.  The input for this models is the vector of current RCA values across all products for the country in question. 
		This approach allows us to create a product progression network, offering insights into potential paths of economic diversification and complexity growth in automotive products for EU countries based on their current export profiles and the relationships between products in the global economic landscape.
		
		\subsection{Data sources}\label{ssec:autom_prods}
		\subsubsection{Product export and automotive supply chain product list}
		Our primary data source for applying the EC framework is the UN-COMTRADE database\footnote{https://comtradeplus.un.org/}, which offers comprehensive information on annual trade flows between countries across more than 5000 categories of goods. Each trade flow is typically reported twice: once by the importing country and once by the exporting country. Due to discrepancies between these declarations, we employ a Bayesian reconciliation procedure to ensure accuracy and consistency in our analysis \citep{patellie2024conomic}.
		
		Goods within the database are classified according to the Harmonized System (HS), which provides a standardized framework for categorizing traded commodities. We rely mostly on the 2007 version of the HS, which serves as the reference standard for our analysis. We then consider HS-2017 for further focus on electric vehicles and batteries since the related categories of products are absent in the previous versions of the classification. Our Economic Complexity analysis is conducted at the finest level of aggregation, namely the 6-digit code level, unless otherwise specified. 
		
		As mentioned above, a crucial methodological step for our analysis is the definition of the products within the automotive supply chain. To propose a novel more comprehensive catalogue of automotive products, we build upon and integrate existing lists in the automotive literature (based on different product classifications)\citep{amighini2014international,blazquez2016international,jetin2018production,klier2006us,turkcan2009vertical,turkcan2011vertical}, primarily expanding on the list proposed by \citet{amighini2014international} in SITC-rev3 product classification.
		
		To enhance this foundation we first convert it to HS 2007 6-digit codes for greater granularity. Second, we refine the list with an empirical validation based on the Istat “Commercio con l’Estero (COE)” database, which provides product-level (HS 2007) imports/exports for the population of Italian firms. To this aim, we select the main 150 inputs imported by firms belonging to the automotive sector based on  specialisation and count share indicators. 
		Finally, cross-referencing this product list based on firm-level information with the lists available in the literature yields a final list of 63 products, belonging not only to the automotive sector but also to rubber and plastics, electrical equipment and other industries. We exclude most basic and fabricated metal inputs due to their standard use as components across multiple industries, not only for motor vehicles manufacture.
		Furthermore, to ensure currency, we convert our product list from HS-2007 to HS-2012 and HS-2017, incorporating new categories included in HS-2007, such as hybrid and electric cars and lithium batteries. However, we note that our analysis does not extend to the full supply chain of electric accumulators or critical raw materials for electric mobility.\footnote{We do not take into account the supply chain of electric accumulators (e.g., lithium-ion accumulators), that is, for instance, we do not measure the graphite flows from its source country to the country where it is processed. A comprehensive analysis of global value chains for critical raw materials and battery components in electric mobility transition is beyond our current scope\citep[see, e.g., ][]{barman2023electric,coffin2018supply,coffin2024electrifying,de2023mapping,lusty2022study,tsuji2022global}}
		The complete automotive product list can be found in the Appendix (Tables \ref{hs07}, \ref{hs12} and \ref{hs17}3).
		
		\subsubsection{Input-Output inter-sectoral linkages}
		We complement the product-level analysis with information on inter-sectoral linkages between the European automotive sector with intra- and extra-EU economies. We employ the OECD Inter-Country Input-Output (ICIO) Tables, which include 76 countries (+ Rest of the World aggregate) and 45 sectors, from 1995 to 2020.\footnote{http://oe.cd/icio.} 
		In our analysis, we consider an aggregated EU27 automotive sector and study its linkages with intra-EU and extra-EU economic branches.
		
		\section{Empirical analysis}\label{sec:results}
		\subsection{The EU27 automotive input dependence}\label{sec:results_input_Dep}
\begin{figure}[htb]
			\centering
			\includegraphics[scale=0.45,trim={0cm 0cm 0cm 0cm}]{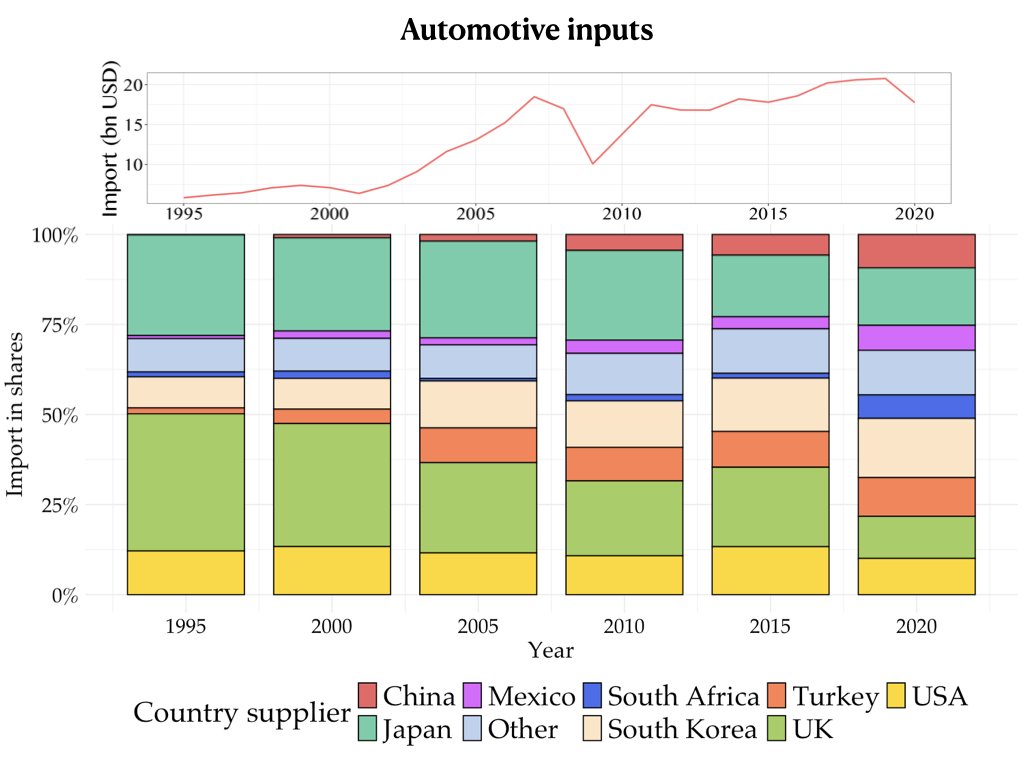}
			\caption{Import in shares of automotive inputs of EU27’s automotive from main Extra-EU27 countries suppliers (1995-2020)}
			\label{io_auto}
		\end{figure}

\begin{figure}[htb]
			\centering
			\includegraphics[scale=0.45,trim={0cm 0cm 0cm 0cm}]{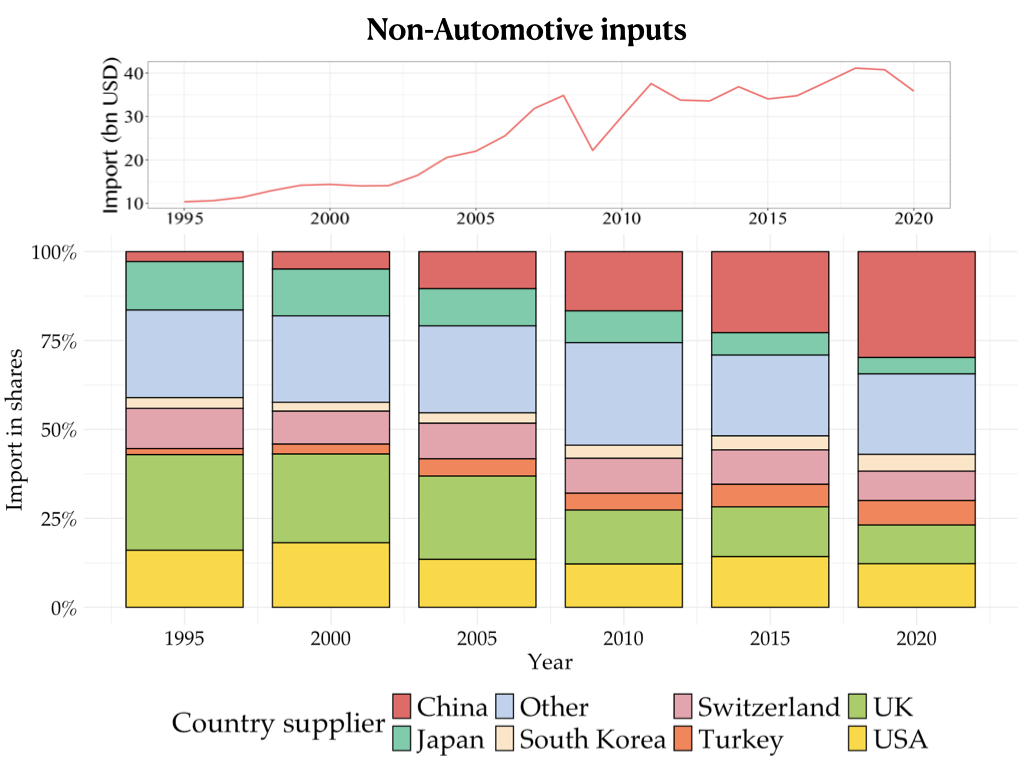}
			\caption{Import in shares of non-automotive inputs of EU27’s automotive from main Extra-EU27 countries suppliers (1995-2020)}
			\label{io_nonauto}
		\end{figure}

		To investigate the possible patterns of vulnerability of European motor vehicle producers,  we show the dependence on inputs sourced from non-EU27 countries, highlighting various relevant dimensions. 
		First, we leverage  Input-Output tables to show the main country and sector suppliers of the EU27 aggregated automotive sector. Figure \ref{io_auto} describes the changing relevance across time (1995-2020) of the main suppliers of inputs coming from Extra-EU automotive sectors.\footnote{Main suppliers are selected by means of a 5\% threshold over EU27 total amount of inputs (i.e., both from within and outside EU27).} Figure \ref{io_nonauto} reports the same information for inputs coming from economic branches other than automotive -- e.g., Rubber and plastics, Basic metals, Electrical equipment.

		From this initial evidence it is possible to observe that the  United Kingdom and Japan are losing  importance as suppliers of EU27 automotive in both automotive and non-automotive goods, in favour of countries like Turkey, South Korea and Mexico, especially for automotive goods. The United States' share for both kinds of inputs is slightly reduced, while China displays the largest share increase from 1995 to 2020, especially in non-automotive inputs.\footnote{Considering only the redistribution of shares between extra-EU27 suppliers.} We also register the role of non-automotive inputs coming from Switzerland with a small but constant share over the whole period under observation.

		Above the two bar plots, we display the time trends in levels for total import in automotive e non-automotive inputs, highlighting that imports from extra-EU27 countries steadily increased from 1995 until the financial crisis of 2007-2008, when a significant decline occurred. Between 2012 and 2015, the trend stabilised, followed by a modest rise until the onset of the COVID-19 pandemics in 2020, which triggered another decline. This pattern suggests a restructuring of the EU automotive supply chain shifting away  from traditional partners such as  the UK and Japan towards China and other suppliers like South Korea, Turkey and Mexico.

\begin{figure}[h]
			\centering
			\includegraphics[scale=0.48,trim={0cm 0cm 0cm 0cm}]{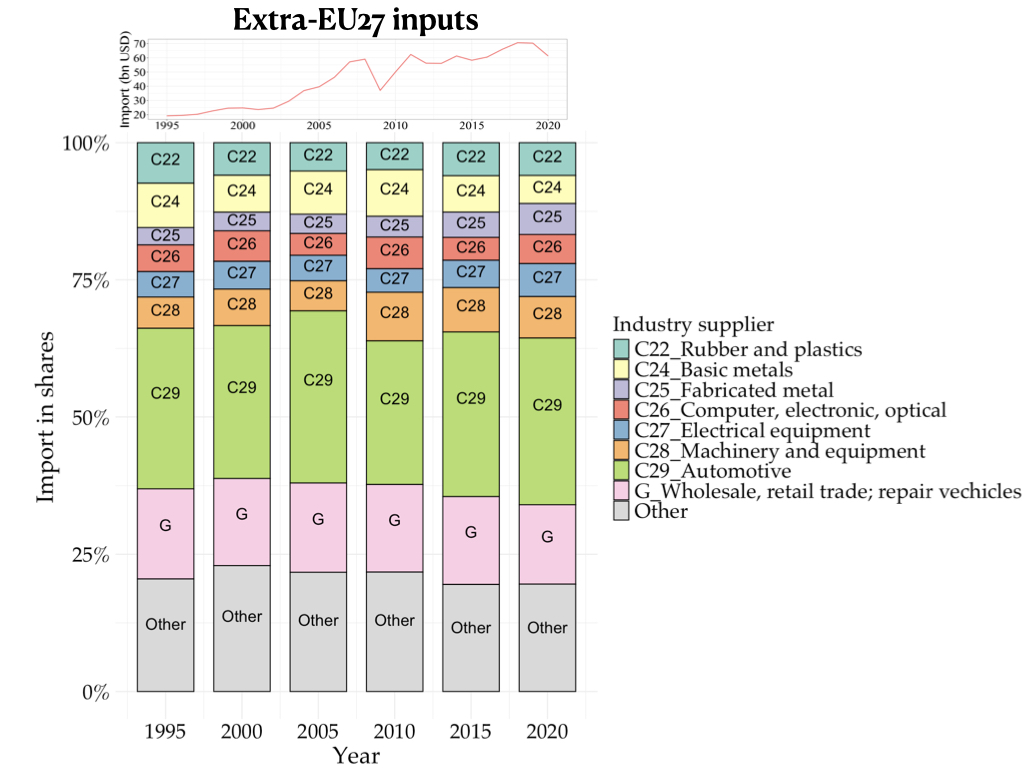}
			\caption{Import in shares of inputs of EU27’s automotive from main industries suppliers of Extra-EU27 countries (1995-2020) }
			\label{io_sector}
		\end{figure}

		Figure \ref{io_sector} instead focuses on the sectoral dimension of the import of inputs for the EU27 automotive industry.\footnote{Main sectoral inputs have been selected with a threshold at 5\% of total extra-EU27 import.} In this case, the shares remain stable across time, with automotive inputs (code C29) being the most relevant sector in the supply chain, indicating that the majority of parts and components needed for motor vehicles production come from the automotive sector. Notably, manufacturing industries such as rubber and plastics, basic metals, electrical equipment, and machinery also play an important role. I-O tables allow us also to account for the considerable relevance of services, including wholesale, retail trade and repair vehicles (code G).
		
		As mentioned above, while I-O tables are valuable for identifying sectoral interdependencies,  they provide information only at the 2-digit industry level, limiting our ability to capture the relevance of more granular intersectoral flows. For instance, within the automotive sector, there are several products that vary widely in terms of technological content and capabilities.
		To gain more granularity and overcome the I-O table strict 2-digit sectoral aggregation, we exploit  UN-COMTRADE 
		and investigate the import of products belonging to the automotive supply chain. We measure the import of 63 intermediate products related to the production of motor vehicles -- see Methods and Data section -- from 2007 to 2022 and highlight the main supplier countries.
		
\begin{figure}[htb]
			\centering
			\includegraphics[scale=0.27,trim={0cm 0cm 0cm 0cm}]{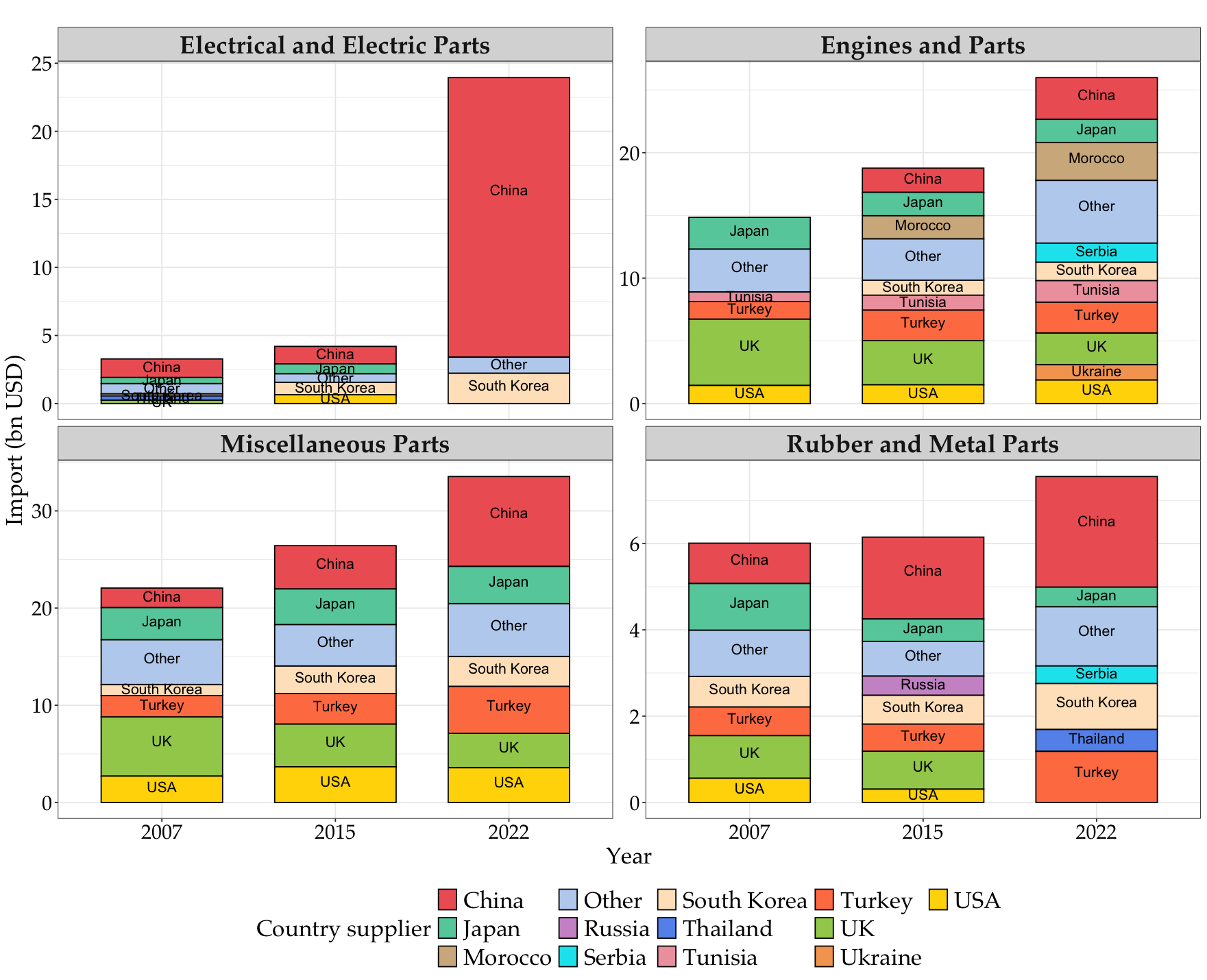}
			\caption{EU27 import of 6-digit HS-2007 products related to the automotive supply chain and aggregated in four categories, highlighting the main suppliers.}
			\label{comtrade_category}
		\end{figure}

		We start by presenting (Figure \ref{comtrade_category}) the import of automotive-related products aggregated by four categories as proposed in \cite{amighini2014international}: Electrical and Electric Parts, Engines and Parts, Miscellaneous Parts, Rubber and Metal Parts. For each category, there is a specific bar plot highlighting the import in billions of USD by main suppliers in 2007, 2015, and 2022.\footnote{Main suppliers are selected based on a threshold at 5\% over total EU27 import from extra EU27 economies. All countries with a share below 5\% are included as ‘Other’.} For each year we highlight the main suppliers and classify the remaining countries in the unique group “Other”.

		The figure highlights a sharp rise in imports  of Electrical and Electric Parts. growing from around \$3.5 billion in 2007 to almost \$25 billion by 2022, with China emerging as the primary supplier, exporting to approximately \$20 billion worth of these goods to the EU27. A similar, though less pronounced, trend can be observed for Miscellaneous Parts, where import increased from around  \$21 billion to over  \$32 billion.
		In this case, China gained a larger market share, the UK lost prominence as supplier, while Turkey and South Korea improved their positions, and Japan and the USA maintained steady shares. Rubber and Metal Parts imports show lower volumes, rising modestly from \$6 to around \$7.5 billion. However, here too, China, Turkey and South Korea's roles have grown significantly. Finally,  imports of Engines and Parts also show an upward trend, from \$15 billion to over \$25 billion, with a larger group of countries entering the supply chain over time, such as Morocco, Serbia, Tunisia, and Ukraine.

\begin{figure}[htb]
			\centering
			\includegraphics[scale=0.45,trim={0cm 0cm 0cm 0cm}]{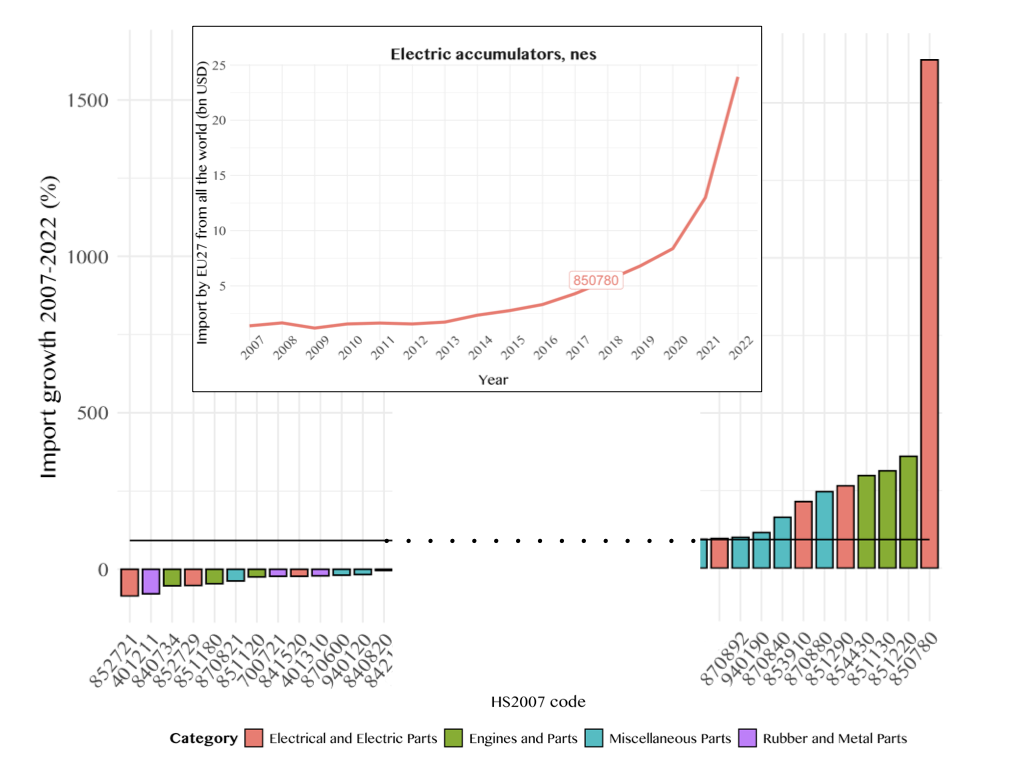}
			\caption{EU27 import growth (\% variation 2007-2022) of 6-digit HS-2007 products related to the automotive supply chain, highlighted by category. The horizontal black line represents the average growth (91.8\% in 2007-2022) of imports for all the products imported by EU27. The line chart represents the time trend over the 2007-2022 period of the product most growing in the import: electric accumulators (code 850780).}
			\label{comtrade_growth}
		\end{figure}

		It is possible to further appreciate the granularity of the information, by plotting the 63 automotive products ranked by import growth between 2007 and 2022, as shown in Figure \ref{comtrade_growth}, where the black horizontal line represents the average import growth for all products imported by EU27.
		
		From the figure it is possible to observe how the import growth of accumulators (code 850780)  outperforms all others – to better appreciate this dynamics, the time trend of electric accumulators imports is superimposed onto the figure – and this may also explain the considerable rise in the import of Electrical and Electric Parts in 2022 observed in Figure \ref{comtrade_category}. 
		Other products for which we register remarkable above-average growth are the following: Lighting/visuals signalling equipment (code 851220), Distributors and ignition coils (code 851130), Ignition/other wiring sets for vehicles/aircraft/ship (code 854430), Parts of cycle and vehicle light, signal, etc equipment (code 851290), Shock absorbers for motor vehicles (code 870880), Transmissions for motor vehicles (code 870840). By emphasising the category the products belong to, it is possible to notice that Rubber and Metal Parts are the components that present the lowest  percentage variation change, while the main changes are observed for Engines and Parts and Electrical and Electric Parts, hinting at a transformation in the EU’s automotive supply chain of these components. Negative but small variations concern mainly products like Radio receivers, external power, sound reproduce/record (code 852721), Retreaded pneumatic tyres of rubber, of a kind used on motor cars (incl. station wagons \& racing cars) (code 401211), Engines, spark-ignition reciprocating, over 1000 cc (code 840734), Radio receivers, external power, not sound reproducer (code 852729).

		\FloatBarrier
		\subsection{EU27’s competitiveness in automotive-related products}\label{sec:results_autom_products}
		By looking at the export dimension and taking advantage of the Economic Fitness and Complexity toolbox, we can shed light on the competitiveness of EU27 in automotive components and vehicles. We compute an automotive sector Fitness of the EU27 member countries using the approach  and the list of 63 products composing the automotive supply chain described in the Methods: for each country we sum the complexity scores of all the automotive products in which the country has a comparative advantage. 
		
\begin{figure}[htb]
			\centering
			\includegraphics[scale=0.35,trim={0cm 0cm 0cm 0cm}]{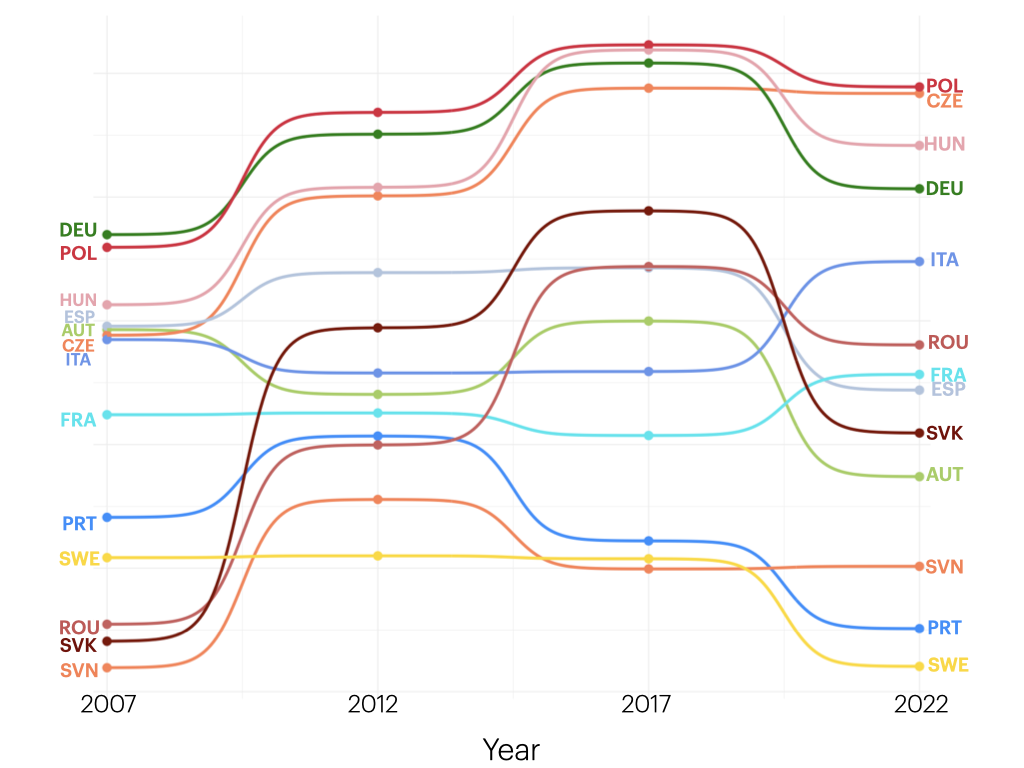}
			\caption{Bump chart of selected EU27 countries for Fitness in 63 automotive inputs in 2007-2022, relativised by maximum value of Fitness.}
			\label{fitness}
		\end{figure}

		In the period from 2007 to 2022, several European Union countries experienced significant shifts in their automotive industry competitiveness, as proxied by the Fitness metric, as shown in the bump chart in Figure \ref{fitness}, where changes in Fitness ranking over time for the most relevant countries, normalised by the maximum Fitness of the automotive supply chain, are displayed.
		
		Austria, Sweden, Spain and Portugal experience a decline in their relative position, while France displays a slight increase in its competitiveness. The most notable changes occurred among Eastern European economies, which register the most positive upward movement in the Fitness ranking. Poland emerges as a leader, gaining the to position by substantially increasing its Fitness and demonstrating considerable specialisation in the automotive sector. Hungary and Czechia also perform increasingly well, both surpassing Germany, with Czechia reaching the top of the ranking. Romania and Slovakia are among the best performers, radically improving their fitness.  Slovenia, though positioned in the middle of the ranking, registers a positive change and overcomes Sweden and Portugal. Italy and Germany are also improving their competitiveness, albeit losing ground with respect to Poland, Czechia and Hungary.\footnote{It must be pointed out that the positive performance of Italy could be explained from a change in specialisation from final to intermediate goods. That is, from 2007 to 2022, Italy has lost ground on the production (and export) of vehicles, turning into a supplier of parts and components for foreign productive units of cars. The supply chain Fitness captures a good performance and capabilities endowments in relation to intermediate products, not on vehicles.}

\begin{figure}[h]
			\centering
			\includegraphics[scale=0.33,trim={0cm 0cm 0cm 0cm}]{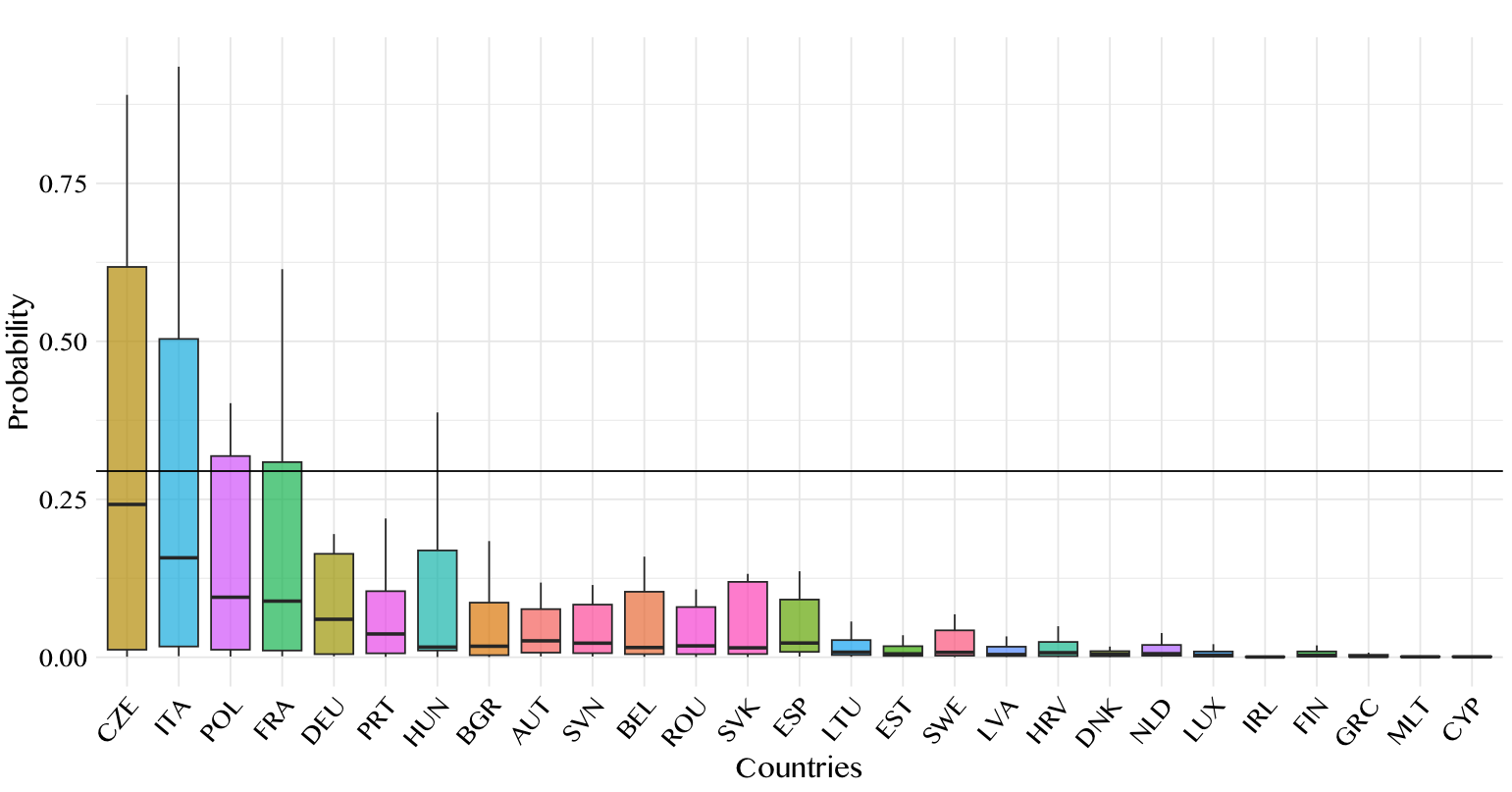}
			\caption{Box-plot of product progression probabilities on the automotive inputs for which each EU27 country does not have a comparative advantage in 2021. 5-year prediction based on 2021. Countries are ranked from left to right based on average probability. The horizontal line represents the average probability on all the 63 automotive-related products at the EU27 level.}
			\label{prog_HS07}
		\end{figure}

		Looking towards the future, the EFC framework can be further used to produce product-level forecasts of countries’ competitiveness through the Product Progression approach described in the Methods. By focusing on the products composing the automotive supply chain, we rank countries based on their average probability of gaining a comparative advantage by 2026 in exporting products for which they did not have a comparative advantage in 2021, as shown in the box-plot in Figure \ref{prog_HS07}, where the horizontal line represents the average probability for all the 63 automotive-related products at the EU27 level. Our findings show that Czechia is leading the ranking, followed closely by Italy, Poland and France. Interestingly, while France lags behind Germany, Italy and some Eastern European countries in 2022 in terms of Fitness, it appears to possess the capabilities to be among the most competitive member states in automotive components in the near future. On the right tail of the distribution are small economies with probabilities below the EU27 average, suggesting they are unlikely to gain significant competitiveness in this supply chain by 2026. 
		
		\FloatBarrier
		
		\subsection{Electric Vehicles}\label{sec:results_EVs}
		
		As discussed in the introduction, the automotive industry has gone through major technological and productive transformations in recent years, most of them related to the transition to electric mobility. In order to investigate the EU27 positioning in the electric mobility race, we exploit the more recent 2017 HS classification, that contains a more fine-grained description of EV-related goods: both a wider range of vehicles (electric and hybrid), and a wider variety of electric accumulators (lithium-ion, lead-acid, nickel-metal and other).\footnote{Electric cars have been more specifically labeled as Battery electric-vehicles (BEV), referring to cars using electric technology as unique form of propulsion, while hybrid vehicles also include an internal combustion engine. The unit of analysis available with UN-COMTRADE data prevent us from further disentangling these two broad categories in more fine grained product types.} 
		
\begin{figure}[htb]
			\centering
			\includegraphics[scale=0.33,trim={0cm 0cm 0cm 0cm}]{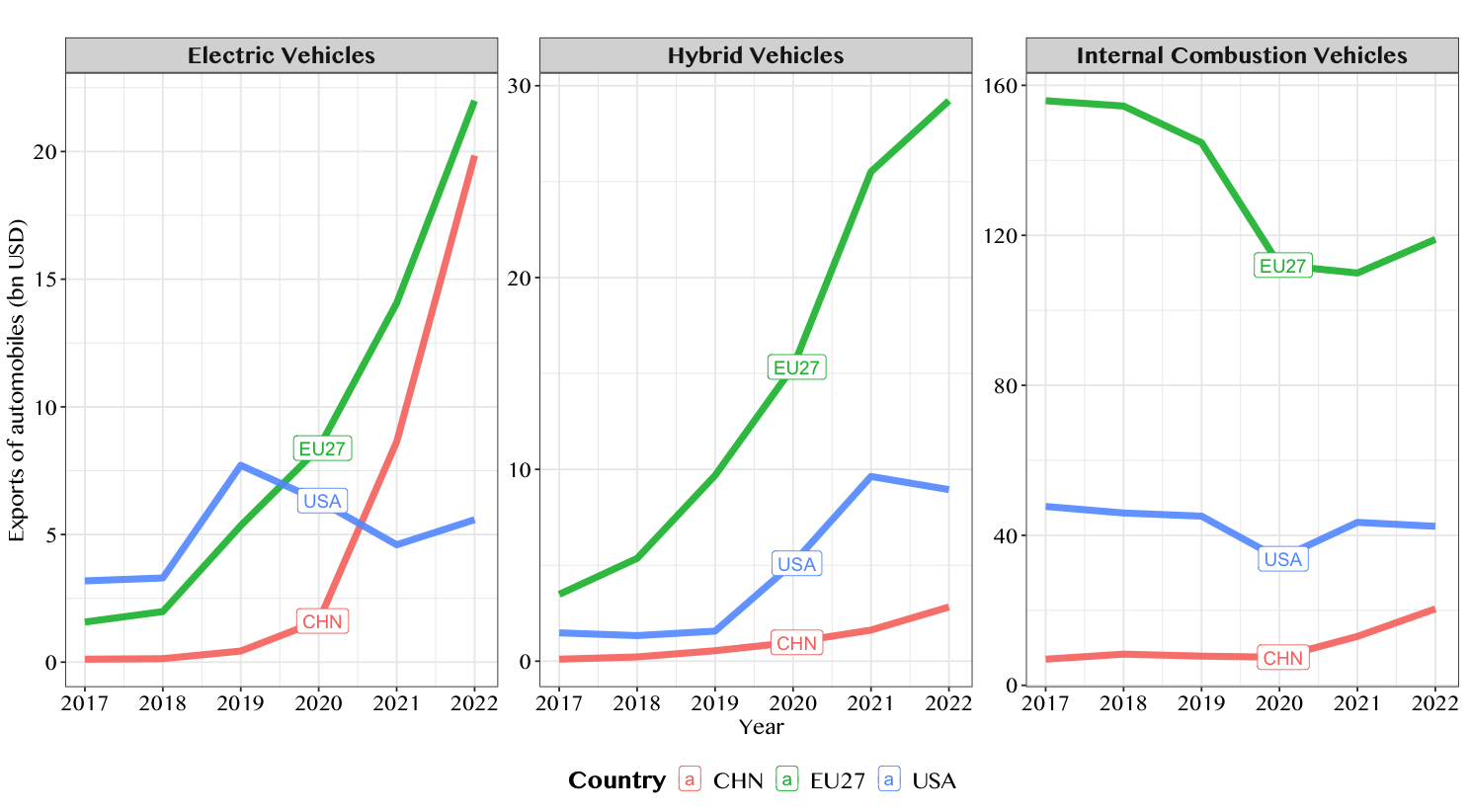}
			\caption{Time trend of the exports of automobiles in HS-2017 comparing  EU27, China and USA in 2017-2021 and distinguished into three main categories: Electric, Hybrid and Internal Combustion vehicles.}
			\label{exp_cars}
		\end{figure}

		In Figure \ref{exp_cars} we present the time trends in exports of three aggregated categories of vehicles -- electric, hybrid and internal combustion engine -- comparing the performance of EU27, China and the US from 2017 to 2022.
		
		For electric and hybrid vehicles, the EU27 shows a significant rise, similar to that of China, reaching around \$22 and \$29 billion, respectively.\footnote{However, it must be noticed that the majority of electric vehicles is exported by Germany, as can be seen in Figure \ref{exp_cars_EU} in the Appendix.} US electric vehicle exports rise sharply until 2019, after which they decline slightly, while hybrid car export remain steady at first, then increase substantially after 2019. China’s export of hybrid vehicles shows a steady increase but remains at the lowest levels compared to the EU and the US. Conversely, China's export of electric vehicles has recently seen a phenomenal rise, surpassing the US in this category. For internal combustion vehicles, the EU27 leads in terms of levels, albeit with a downward trend. Meanwhile, the US and China show more stable, flatter dynamics. Notably,  internal combustion vehicle exports remain significantly higher than those of electric and hybrid cars.

\begin{figure}[h]
			\centering
			\includegraphics[scale=0.33,trim={0cm 0cm 0cm 0cm}]{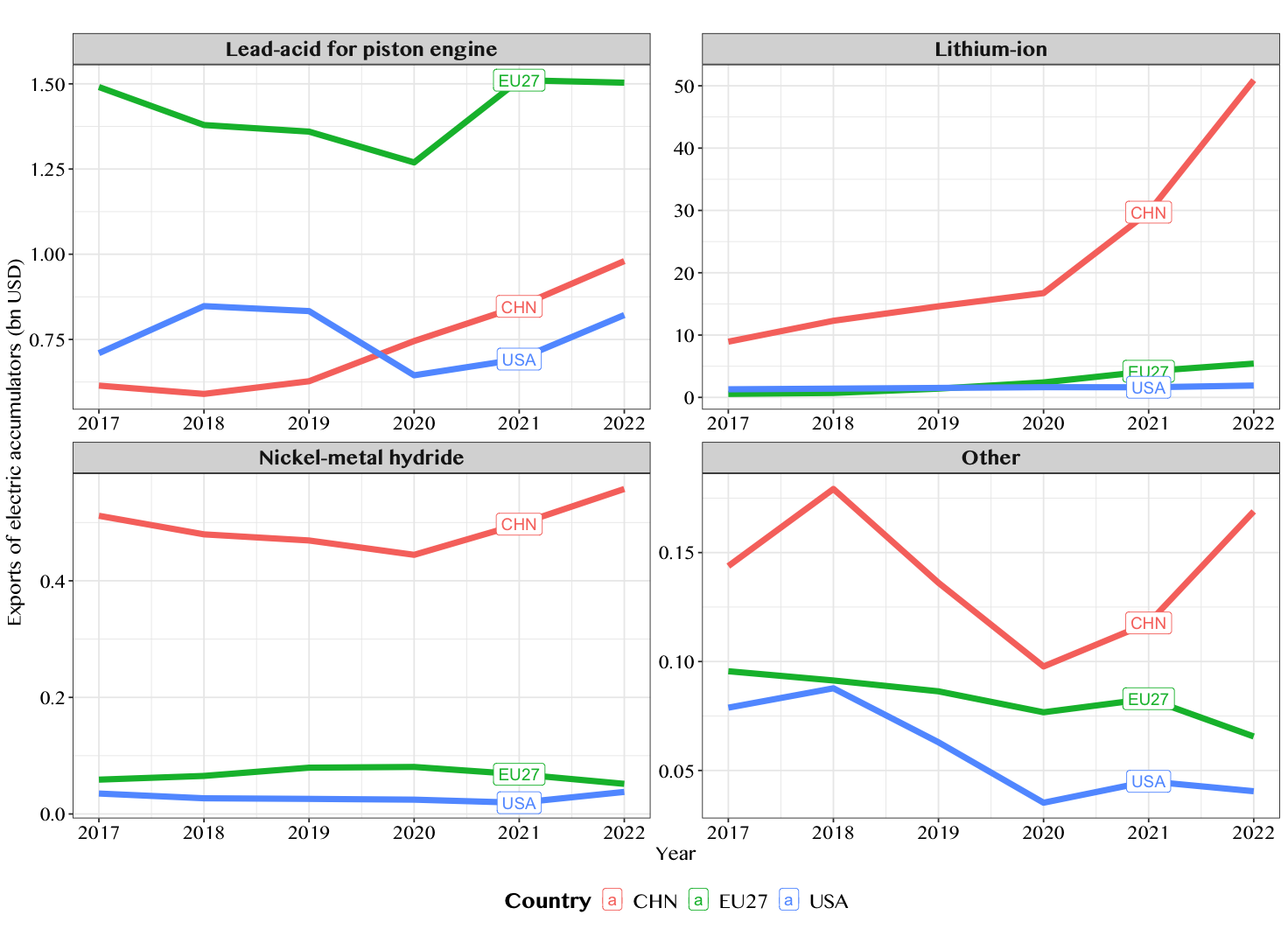}
			\caption{Time trend of the exports of electric accumulators in HS-2017 for aggregated EU27 in 2017-2022 and distinguished for typology: Lead-acid for piston engine, Lithium-ion, Nickel-metal hydride and Other.}
			\label{exp_acc}
		\end{figure}

		A strategic input for the electric mobility transition is the export and import of  electric accumulators, and we already discussed the growth of EU’s accumulator imports. By exploiting the granularity of HS-2017 we analyse the export and the import of four categories of accumulators.\footnote{Lead-acid for piston engines, nickel-metal hydride and lithium-ion electric accumulators are commonly used in motor vehicles, with the lithium-ion batteries being the most prevalent due to their superiorior enrgy accumulation and lower weight. We also include the category “Other”, hough we lack specific details on the battery types it encompasses.} As shown in Figure~\ref{exp_acc}, where again we compare the performances of EU27, China and the US, lithium-ion batteries represent the largest export market, with China leading the way, showing a consistent rise and exceeding \$20 billion.
		
		In contrast, the US and EU27 are lagging behind, with the latter performing slightly better and registering a steady export increase since 2019, nearing \$5 billion. Lead-acid, nickel-metal and other batteries remain below \$1 billion.  The EU27 outperforms China and the US in lead-acid exports, while China is leads in nickel-metal and other batteries, followed by the EU and the US. 	 
		
		This granularity in electric accumulators can also be used to assess the import dimension for the EU27, in line with the analysis performed above. Figure \ref{exp_acc} illustrates the import trends of the three kinds of batteries, with lithium-ion batteries showing the highest import figures and a constant rise throughout the period. Nickel-metal hybrid accumulators remain stable, while lead-acid accumulators  show a slight increase. The "Other batteries" HS products exhibits a notable rise in 2020, remains stable in 2021, and then declines in 2022.

\begin{figure}[h]
			\centering
			\includegraphics[scale=0.28,trim={0cm 0cm 0cm 0cm}]{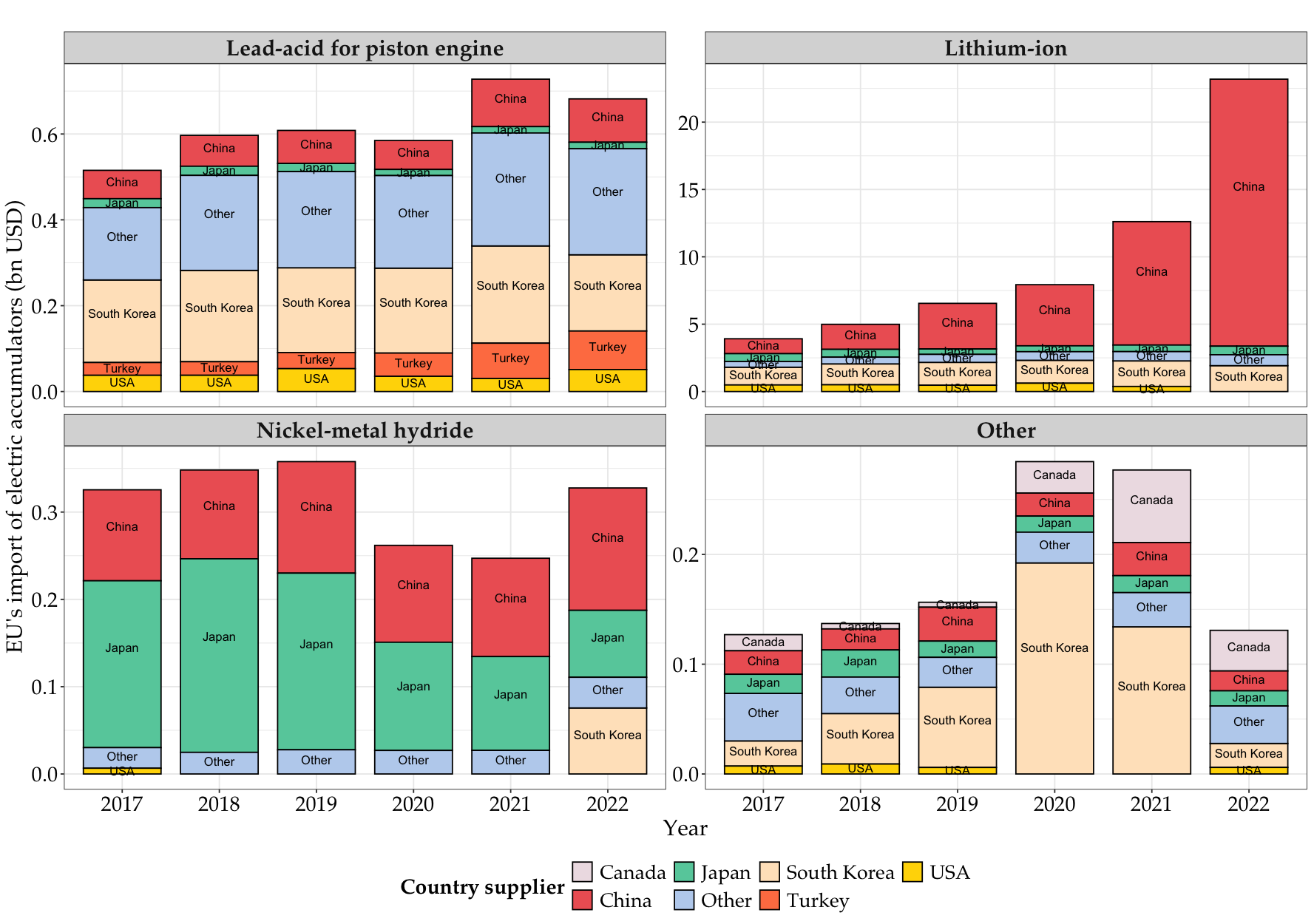}
			\caption{Time trend of the imports of electric accumulators in HS-2017 for aggregated EU27 in 2017-2022 and distinguished for typology and main country supplier.}
			\label{imp_acc}
		\end{figure}

		Figure \ref{imp_acc} further examines  the import of electric accumulators, emphasising the  composition of suppliers from 2017 to 2022.
		The dominant role of China is confirmed, especially for lithium-ion batteries. South Korea  is a significant supplier of lead-acid and other accumulators, while Japan, alongside China,  is a key supplier of nickel-metal hydrid batteries. Other countries emerging as main suppliers are Turkey and the US for lead-acid batteries and Canada for other types of accumulators.\footnote{For the "Other" category, an issue with COMTRADE data was noted. As illustrated in Figure \ref{imp_acc}, the import of these batteries drops significantly in 2022, largely due to Slovakia's imports of Korean electric accumulators falling from 130 million USD in 2021 to 19 million USD in 2022. However, when examining a higher aggregation level (HS2007 code 8507, 4-digit level), the decline is smaller, indicating a potential data reporting error.}
		
\begin{figure}[h]
			\centering
			\includegraphics[scale=0.33,trim={0cm 0cm 0cm 0cm}]{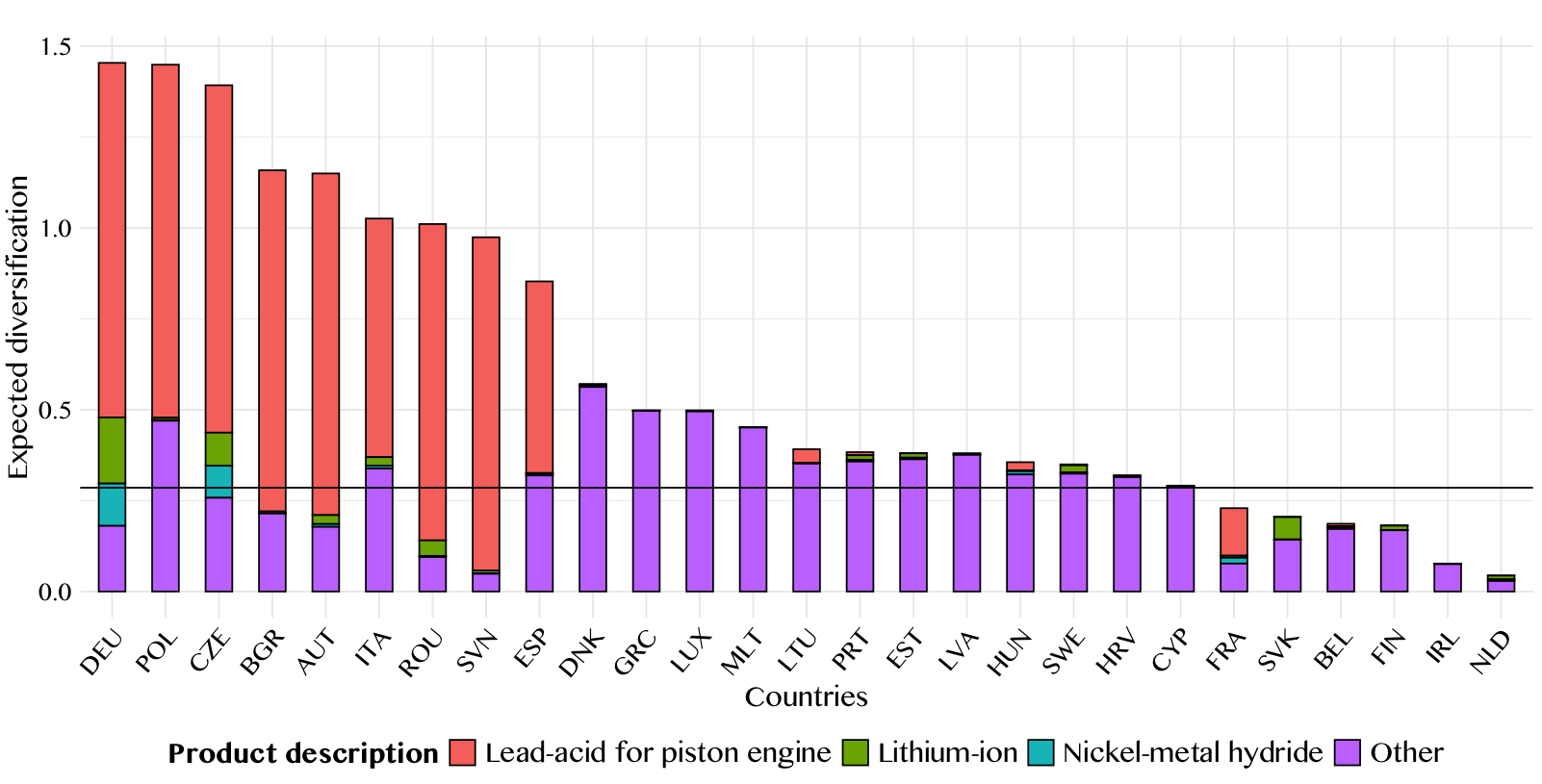}
			\caption{Bar plot of the expected diversification on electric accumulators for each European Member State, given by the product progression probabilities for the four different kinds of accumulators. The probability is a 5-year prediction based on 2022 HS-2012 data. The horizontal line represents the average probability on automotive inputs at the EU27 level.}
			\label{prog_acc}
		\end{figure}

		Finally, reverting to the HS 2012 classification, we compute the product progression probability for four products related to electric accumulators: lead-acid for piston engines, lithium-ion, nickel-metal hydrid, and other. As shown in Figure \ref{prog_acc}, EU27 member states are more likely to produce two main categories: lead-acid piston engine batteries and those not included in the three main categories. The probability of exporting lithium accumulators appears is small and limited to a few countries, such as Germany, Czechia, Romania and Slovakia. Germany and Czechia are the only economies showing a non-zero probability of exporting nickel-metal hydrid accumulators. n terms of the overall ranking across all four battery types, Germany stands at the top, followed by Poland and Czechia, as the countries with highest total probabilities. In contrast, at the bottom of the ranking, we find France and Finland, alongside with smaller economies such as Slovakia, Belgium, Ireland and the Netherlands.

		\FloatBarrier
		\subsection{Vulnerability analysis}\label{sec:results_vulnerability}
		The current times have been marked by geopolitical tensions, pandemics, natural disasters and structural changes driven by the green transition. These phenomena often manifest as shocks primarily affecting the supply chains of various manufacturing commodities. Considerable concern has emerged among policy makers regarding the vulnerabilities of productive systems, which are the result of decades of increasing fragmentation of production, the so-called "age of global value chains". This phase is now giving way to de-globalization, characterised by reshoring, friendshoring, and a general shortening of these productive chains \citep{maihold2022new,banaszyk2023reshoring,alfaro2023global,pisani2024risk}. 
		In response,  the academic literature has proposed various methods to identify vulnerabilities within global production networks. A detailed review of recent contributions can be found in \cite[][see p.16, Box 2.1]{berthou2024mapping}. The European Commission itself \citep{EC2021} has proposed an approach to define vulnerable products by relying on import concentration and substitutability measures. \\
		
		In this section, we perform a vulnerability analysis of the European automotive supply chain, building upon this research stream by constructing two indicators at the HS2012 6-digit product level:\footnote{We explot the HS2012 classification because it includes a more detailed representation of electric accumulators.}
		\begin{itemize}
			\item A measure of net exposure to extra-EU27 automotive inputs, computed as the ratio of imports from extra-EU27 to imports from within the EU27. 
			\item An Herfindal-Hirschman concentration index (HHI-M) for product-level imports, in the spirit of \cite{berthou2024mapping}'s "supply chain concentration" index,, defined as the sum of the square shares of the supplier countries:
			\begin{equation}
				\text{HHI-M}_{i} = \sum\limits_{h=1}^{H} s_{i,j}^{2}
			\end{equation}
			Where for each automotive input $i$:
			\begin{equation}
				s_{i,j} = \dfrac{M_{i,j}}{M_{j}}
			\end{equation}
			Where $M_{i,j}$ is the import of product $i$ from extra-EU27 country $j$, and $M_{}j$ is the total import of automotive input from extra-EU27 countries. 
		\end{itemize}
		
\begin{figure}[h]
			\centering
			\includegraphics[scale=0.5,trim={1cm 0cm 0cm 0cm}]{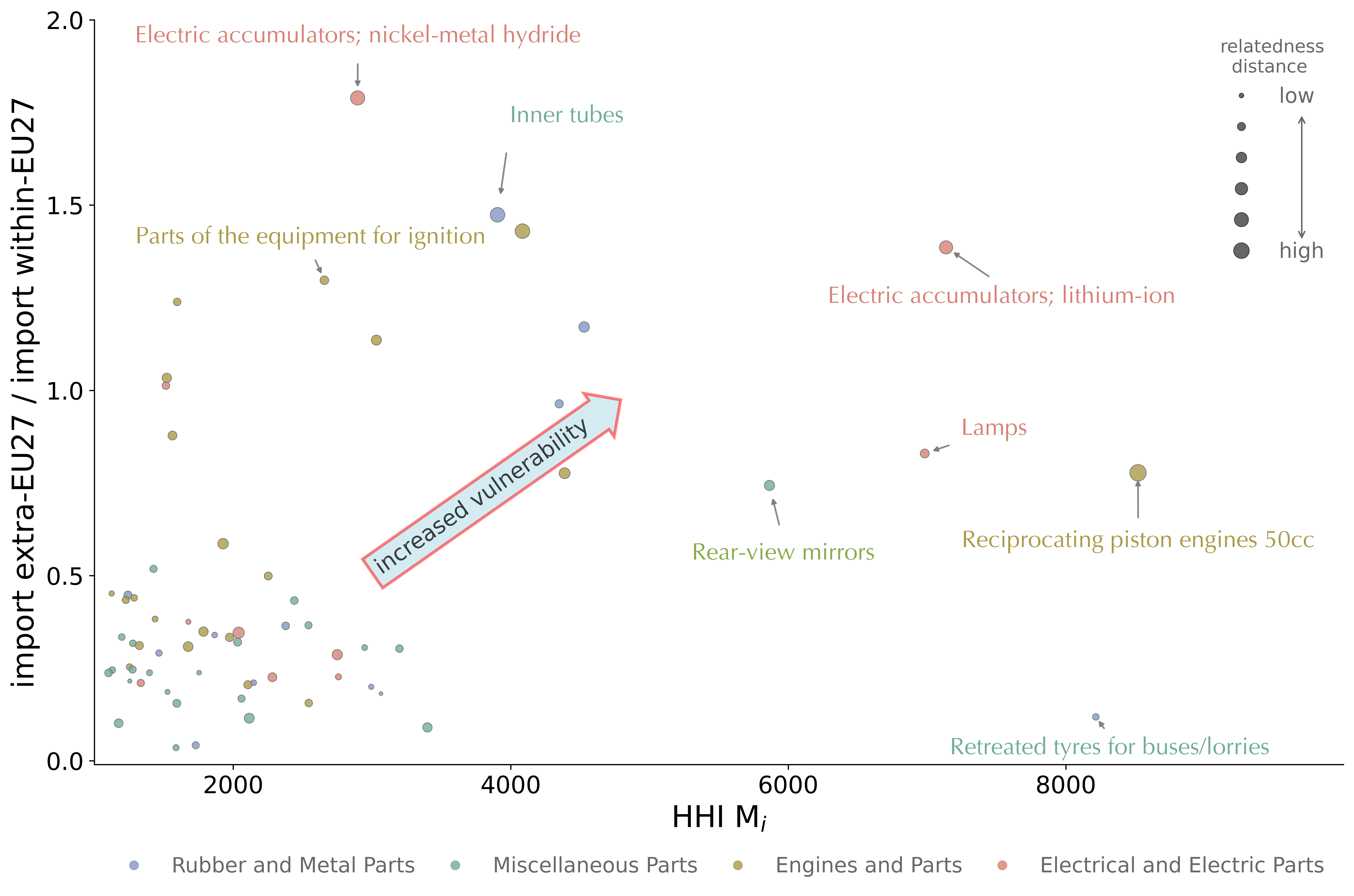}
			\caption{Scatter plot of the 65 automotive inputs in 2022. On the y-axis the ratio between import from extra-EU27 and import from within EU27, on the x-axis the supply chain concentration index at the EU level. Each point's size is proportional to the "relatedness distance", captured by the log of the inverse of the 5-year product progression probability in 2022.}
			\label{vulnerabilty}
		\end{figure}

		In the scatter plot in Figure \ref{vulnerabilty}, products are distributed based on their vulnerability, for 2022, with the two indicators on the axes. The y-axis represents the ratio of imports from extra-EU27 countries compared to imports from within the EU27, while the x-axis shows the supply chain concentration index, which reflects the level of dependence on a small number of suppliers. The size of each point, representing a product, depends on the logarithm of the inverse of the product progression probability, meaning larger points indicate a lower likelihood of exporting that product in five years. Each point is colour-coded according to the macro product category which it belongs to.
		
		Products located in the lower-left area of the plot tend to have both low exposure to extra-EU27 inputs and low supply chain concentration, indicating a diversified and resilient supply chain with multiple suppliers. These products are less vulnerable to external shocks. In contrast, products in the upper-right area of the plot exhibit higher levels of exposure to extra-EU27 inputs and higher supply chain concentration, meaning they are primarily sourced from a few countries outside the EU27, making them more susceptible to disruptions. Electric accumulators, for example, stand out with a high HHI-M concentration index and significant vulnerability, largely due to dependence on imports from China. The relatively large size of the marker for electric accumulators indicates a lower expected capability to export this product in the future, which is tied to the scarcity of critical raw materials like lithium within the EU27.
		Looking at the macro product categories, those with the least vulnerability are often from the Miscellaneous Parts category (yellow points), while several products from the Electrical and Electric Parts (red) and Engines and Parts (blue) categories are more exposed, with a skewed import ratio toward extra-EU27 sources. These categories encompass many of the more technologically advanced goods in the automotive supply chain.

		\FloatBarrier
		\section{Conclusion}\label{sec:conclusion}
		This paper leverages on two complementary approaches -- Input-Output Analysis and Economic Fitness and Complexity -- to analyse the vulnerability and the competitiveness of the European automotive industry. We applied these methods using two key data sources and providing insights at the sector- and product-level: OECD Inter-Country Input-Output Tables, focusing on 2-digit sectoral linkages, and UN-COMTRADE data, focusing on the 6-digit product level to provide a granular examination of automotive supply chains. The main methodological contribution of our paper is the identification of the 6-digit products belonging to the motor vehicles' supply chain, by refining the existing product lists by validating them with firm-level data and converting them into the Harmonised System classifications for 2007, 2012 and 2017. 
		
		Through I-O analysis and mixing information from I-O tables and product-level import data, we measured the EU27 automotive industry's growing dependency on Extra-EU27 inputs. China has emerged as the dominant supplier for electrical equipments and especially for lithium-ion batteries, which are becoming the main power source for EVs and are experiencing a sky-rocketing import increase. The shift in suppliers also highlights the rising importance of countries like Mexico, Turkey, and Morocco in the EU’s automotive supply chain, while Japan and South Korea, although still significant, saw a relative decline. The product-level analysis underscores that the main transformations concern electrical and electric parts, with electric accumulators showing the highest import growth within the EU27, confirming the rising vulnerability in this segment. Our vulnerability analysis confirms the risks associated with the import goods belonging to the categories of Electrical and Electric Parts or Engines and Parts. This is particularly true for electric accumulators, which are primarily imported from a few extra-EU27 suppliers, with China being the dominant player. The concentration of supply and the EU’s limited battery production capabilities in these critical components raise concerns about the region’s ability to secure a stable supply of essential inputs, especially as the demand for EV-related products grows rapidly.
		
		In terms of competitiveness, the EFC analysis provides insights into the export specialization of EU27 countries in the automotive sector.  Our findings reveal a considerable increase in EU27 exports of battery electric vehicles and of hybrid vehicles, although internal combustion engine vehicles still dominate the EU’s automotive export mix. When looking at the Fitness of EU27 countries in the export of automotive components, that takes into account their trade specialization and the complexity of their exports, our analysis shows that Eastern European countries, particularly Poland, Hungary, and Czechia, have significantly improved their competitiveness in automotive components, surpassing Germany and Italy. 
		The product progression probability analysis shows also that these countries also display the highest probabilities of developing future comparative advantages in automotive-related products over the next five years. Czechia, followed by Italy, displays the highest 5-year probability of becoming specialised in exporting the parts and components for which it has not a comparative advantage in 2021. The EU27 is mainly specialised in lead-acid batteries for piston engine accumulators and in other minor categories, but still lags behind in the production and export of lithium-ion batteries, with only a few countries, such as Germany, Czechia and Slovakia, showing modest chances of becoming competitive in this market by 2027. While the rise of Eastern European countries in the automotive supply chain suggests that they could play a key role in supporting the EU’s industrial transition, particularly in the production of intermediate goods, this would require targeted investments in infrastructure and capabilities to ensure that these countries can continue to upgrade their position within the automotive value chain.
		
		The broader implications of our findings suggest that the European automotive industry remains strategically important, with some countries especially increasingly specialised in automotive and others displaying potential for increasing their competitiveness, but faces increasing risks due to external dependencies and supply chain vulnerabilities. In fact, we register the rising dependence of the EU27 from countries like Turkey and Morocco -- where parts of the European automotive production have been offshored -- but especially the current reliance on electric accumulator imports from China. 
		This presents a significant vulnerability for the EU27 as it seeks to transition to electric mobility and compete in the global EV market because, according to our Product Progression Probability analysis, it is unlikely that the EU27 will become specialised in the production and export of accumulators, as probably due to the dependence on external sources for critical raw materials, as well as insufficient domestic capabilities and productive capacity for processing these materials and assembling batteries.
		Recent policy proposals by the European Commission, such as tariffs on China-made electric cars,\footnote{\url{https://ec.europa.eu/commission/presscorner/detail/en/ip_24_3630}} provide some response, however there is still insufficient focus on interventions aimed at preserving existing manufacturing capabilities while simultaneously developing new ones to support the electric mobility transition. Although the debate on industrial policy at the EU level has been revitalized in light of recent supply chain shocks from the pandemic and geopolitical upheaval \citep{coveri2020supply,crespi2021european,pichler2021eu,klebaner2022european,di2023europe,megyeri2023realities}, more concrete actions are needed. 
		
		
		
		Mario Draghi's 2024 report on European competitiveness advocates for stronger vertical and horizontal coordination across value chains in the automotive sector, 
		albeit pointing to a problematic detachment between EU climate policies, which set ambitious targets for low-carbon mobility and cleaner vehicles production,\footnote{The European Green Deal targets climate neutrality by 2050 (for more information see the dedicated website of the European Commission: \url{https://climate.ec.europa.eu/eu-action/climate-strategies-targets/2050-long-term-strategy_it}).} and the EU's industrial policies, , which have been slow to adapt to these transitions.
		This disconnect is compounded by the lack of coordination and long-term strategic planning at EU level. In contrast, the US implemented a large-scale support for domestic value chains through the Inflation Reduction Act, along with protective trade measure against China, which have significantly lowered its production costs and gained a technological edge after years of massive investments in all industries related to the EV lifecycle, including the sourcing and processing of critical raw materials. 
		At the EU level, several initiatives have been introduced to address these gaps, though many lack adequate financial backing. Key measures include the European Battery Alliance (2017), aimed at reducing dependency and building a sustainable battery value chain within Europe, and the European Chips Act (2022), which seeks to achieve a 20\% global share of semiconductor production by 2030.\footnote{As the automotive industry becomes more and more reliant on semiconductors for electric and autonomous vehicle production, securing a steady supply is crucial for Europe's industrial future.} The Critical Raw Materials Act (2023) focuses on securing essential materials like lithium, cobalt, and nickel while promoting domestic mining, recycling, and processing capacity, while the Net Zero Industry Act (2023)'s goal is to advance the EU’s green transition by reducing reliance on external sources for clean technologies. Additionally, the Important Projects of Common European Interest (IPCEI) fosters collaboration between the European Commission and national governments to develop strategic markets and supply chains in sectors such as raw materials, batteries, hydrogen, semiconductors, and cutting-edge technologies. However, even within these existing EU policies, questions remain regarding who will implement these interventions, the scale of investments required, and the sources of funding.

		This work opens up different future research avenues that may enrich our understanding of the European automotive industry and its supply chain. First, future research may extend this analysis by examining the complete value chain of electric accumulators, from the sourcing and processing of critical raw materials to the assembling of batteries. Second, firm-level analysis, taking into account the main European cars manufacturers, would help clarify whether the EU’s dependence on extra-EU27 inputs is linked to the Extra-EU27 establishments where European car manufacturers offshore production. This may be informative when assessing the feasibility of reshoring, near-shoring or friend-shoring in the current phase of de-risking and supply chain shortening \citep{maihold2022new,banaszyk2023reshoring,alfaro2023global,pisani2024risk}. 
		Finally, further exploration of the emergence of Eastern European countries, such as Poland, Czechia, and Hungary -- evidenced by their considerable rise in Fitness in the export of automotive components -- could provide valuable insights into the dynamics of vertical integration and supply chain management by examining the structural interdependencies between these economies and core EU countries like Germany. A possible research direction could involve developing a way to weigh the Supply Chain Fitness indicator by accounting for vertical integration, where an increase in export specialization in emerging economies might reflect significant offshoring from mature economies, which continue to dominate as the heads of the supply chain. These power relationships and structural dependencies represent interesting dimensions to be add to a broader analysis taking advantage of the fruitful bridge between I-O analysis and EFC methods we have proposed.
		\section*{Acknowledgements}
		We wish to thank the participants of the following conferences for the fruitful comments and suggestions provided: \textit{Conference on Complex Systems} (October 2023, Salvador, Brazil), \textit{7th International Astril Conference} (January 2024, Rome, Italy), \textit{Catching up and global value chains in times of transformation - Final CatChain Symposium} (February 2024, Maastricht, The Netherlands), \textit{Capabilities, GVCs and productive complexity: Comparing frameworks to explore the green transition and supply chain vulnerability} (April 2024, Rome, Italy), The 32nd International Colloquium of Gerpisa - Convergence and divergence of trajectories of change, (June 2024, Bordeaux, France), \textit{XLV Conferenza Scientifica annuale AISRe} (September 2024, Turin, Italy).
		
		D.M., A.P., A.S., and A.T. acknowledge the financial support under the National Recovery and Resilience Plan (NRRP), Mission 4, Component 2, Investment 1.1, Call for tender No. 1409 published on 14.9.2022 by the Italian Ministry of University and Research (MUR), funded by the European Union – Next Generation EU – Project Title "Triple T – Tackling a just Twin Transition: a complexity approach to the geography of capabilities, labour markets and inequalities" – CUP F53D23010800001 - Grant Assignment Decree No. 1378 adopted on 01.09.2023 by the Italian Ministry of Ministry of University and Research (MUR).
		
		D.M., A.P., A.S., and A.T. acknowledge the financial support under the National Recovery and Resilience Plan (NRRP), Mission 4, Component 2, Investment 1.1, Call for tender No. 104 published on 2.2.2022 by the Italian Ministry of University and Research (MUR), funded by the European Union – NextGenerationEU– Project Title  "WECARE" – CUP 20223W2JKJ by the Italian Ministry of Ministry of University and Research (MUR).
		
		\bibliography{Ref}
		
		\newpage
		\begin{appendices}
			
			\begin{figure}[htb]
				\centering
				\includegraphics[scale=0.33,trim={0cm 0cm 0cm 0cm}]{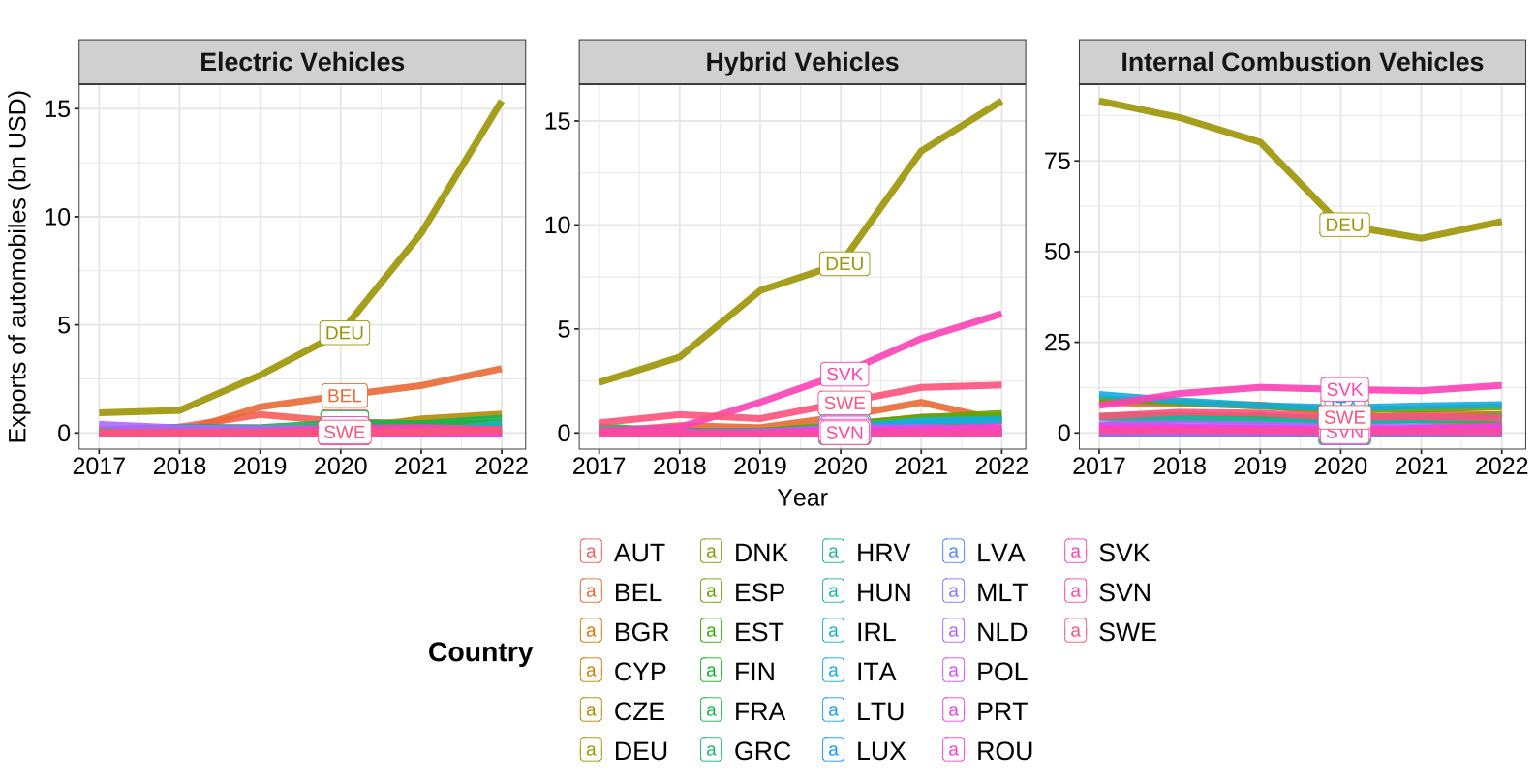}
				\caption{Time trend of the exports of automobiles in HS-2017 comparing EU Member States in 2017-2021 and distinguished into three main categories: Electric, Hybrid and Internal Combustion vehicles.}
				\label{exp_cars_EU}
			\end{figure}

			\begin{longtblr}[
				caption = {List of products in HS 2007},
				label = {up-down},
				]{
					colspec = {|p{5.5cm}|p{1.5cm}|p{9cm}|},
					rowhead = 2,
					hlines,
					row{even} = {gray9},
					row{1} = {CadetBlue!30},
				} 
				
				\hline
				\textbf{Category}   & \textbf{HS07 code} & \textbf{HS07 description} \\
				\hline
				Rubber and Metal Parts: &  401110    &   Pneumatic tyres new of rubber for motor cars    \\
				Rubber and Metal Parts        & 401120              & Pneumatic tyres new of rubber for buses or lorries                                                        \\
				Rubber and Metal Parts        & 401211              & Retreaded pneumatic tyres of rubber, of a kind used on motor cars (incl.   station wagons \& racing cars) \\
				Rubber and Metal Parts        & 401212              & Retreaded pneumatic tyres of rubber, of a kind used on buses/lorries                                      \\
				Rubber and Metal Parts        & 401290              & Solid or cushioned tyres, interchangeable treads                                                          \\
				Rubber and Metal Parts        & 401310              & Inner tubes of rubber for motor vehicles                                                                  \\
				Rubber and Metal Parts        & 700711              & Safety glass (tempered) for vehicles, aircraft, etc                                                       \\
				Rubber and Metal Parts        & 700721              & Safety glass (laminated) for vehicles, aircraft, etc                                                      \\
				Miscellaneous Parts           & 700910              & Rear-view mirrors for vehicles                                                                            \\
				Rubber and Metal Parts        & 731600              & Anchors, grapnels and parts thereof of iron or steel                                                      \\
				Rubber and Metal Parts        & 732010              & Leaf springs/leaves thereof, iron or steel                                                                \\
				Miscellaneous Parts           & 830120              & Locks of a kind used for motor vehicles of base metal                                                     \\
				Rubber and Metal Parts        & 830230              & Motor vehicle mountings, fittings, of base metal, nes                                                     \\
				Engines and Parts             & 840731              & Engines, spark-ignition reciprocating, \textless{}50 cc                                                   \\
				Engines and Parts             & 840732              & Engines, spark-ignition reciprocating, 50-250 cc                                                          \\
				Engines and Parts             & 840733              & Engines, spark-ignition reciprocating, 250-1000 cc                                                        \\
				Engines and Parts             & 840734              & Engines, spark-ignition reciprocating, over 1000 cc                                                       \\
				Engines and Parts             & 840820              & Engines, diesel, for motor vehicles                                                                       \\
				Engines and Parts             & 840991              & Parts for spark-ignition engines except aircraft                                                          \\
				Engines and Parts             & 840999              & Parts for diesel and semi-diesel engines                                                                  \\
				Electrical and Electric Parts & 841520              & Air Conditioning Machines of A Kind Used For Persons In Motor Vehicles                                    \\
				Miscellaneous Parts           & 848310              & Transmission shafts and cranks, cam and crank shafts                                                      \\
				Electrical and Electric Parts & 850710              & Lead-acid electric accumulators (vehicle)                                                                 \\
				Electrical and Electric Parts & 850780              & Electric accumulators, nes                                                                                \\
				Engines and Parts             & 851110              & Spark plugs                                                                                               \\
				Engines and Parts             & 851120              & Ignition magnetos, magneto-generators and flywheels                                                       \\
				Engines and Parts             & 851130              & Distributors and ignition coils                                                                           \\
				Engines and Parts             & 851140              & Starter motors                                                                                            \\
				Engines and Parts             & 851150              & Generators and alternators                                                                                \\
				Engines and Parts             & 851180              & Glow plugs \& other ignition or starting equipment nes                                                    \\
				Engines and Parts             & 851190              & Parts of electrical ignition or starting equipment                                                        \\
				Engines and Parts             & 851220              & Lighting/visual signalling equipment nes                                                                  \\
				Engines and Parts             & 851230              & Sound signalling equipment                                                                                \\
				Engines and Parts             & 851240              & Windscreen wipers/defrosters/demisters                                                                    \\
				Electrical and Electric Parts & 851290              & Parts of cycle \& vehicle light, signal, etc equipment                                                    \\
				Electrical and Electric Parts & 852721              & Radio receivers, external power,sound reproduce/recor                                                     \\
				Electrical and Electric Parts & 852729              & Radio receivers, external power, not sound reproducer                                                     \\
				Electrical and Electric Parts & 853910              & Sealed beam lamp units                                                                                    \\
				Engines and Parts             & 854430              & Ignition/other wiring sets for vehicles/aircraft/ship                                                     \\
				Electrical and Electric Parts & 854810              & Waste \& Scrap of Primary Cells, Primary Batteries and Electric   Accumulators;                           \\
				Vehicles                      & 870321              & Automobiles, spark ignition engine of \textless{}1000 cc                                                  \\
				Vehicles                      & 870322              & Automobiles, spark ignition engine of 1000-1500 cc                                                        \\
				Vehicles                      & 870323              & Automobiles, spark ignition engine of 1500-3000 cc                                                        \\
				Vehicles                      & 870324              & Automobiles, spark ignition engine of \textgreater{}3000 cc                                               \\
				Vehicles                      & 870331              & Automobiles, diesel engine of \textless{}1500 cc                                                          \\
				Vehicles                      & 870332              & Automobiles, diesel engine of 1500-2500 cc                                                                \\
				Vehicles                      & 870333              & Automobiles, diesel engine of \textgreater{}2500 cc                                                       \\
				Vehicles                      & 870390              & Automobiles nes including gas turbine powered                                                             \\
				Miscellaneous Parts           & 870600              & Motor vehicle chassis fitted with engine                                                                  \\
				Miscellaneous Parts           & 870710              & Bodies for passenger carrying vehicles                                                                    \\
				Miscellaneous Parts           & 870790              & Bodies for tractors, buses, trucks etc                                                                    \\
				Miscellaneous Parts           & 870810              & Bumpers and parts thereof for motor vehicles                                                              \\
				Miscellaneous Parts           & 870821              & Safety seat belts for motor vehicles                                                                      \\
				Miscellaneous Parts           & 870829              & Parts and accessories of bodies nes for motor vehicle                                                     \\
				Miscellaneous Parts           & 870830              & Brakes \& servo-brakes; parts thereof, of the motor vehicles of   headings 87.01 to 87.05.                \\
				Miscellaneous Parts           & 870840              & Transmissions for motor vehicles                                                                          \\
				Miscellaneous Parts           & 870850              & Drive axles with differential for motor vehicles                                                          \\
				Miscellaneous Parts           & 870870              & Wheels including parts/accessories for motor vehicles                                                     \\
				Miscellaneous Parts           & 870880              & Shock absorbers for motor vehicles                                                                        \\
				Engines and Parts             & 842123              & Oil/petrol filters for internal combustion engines                                                        \\
				Engines and Parts             & 842131              & Intake air filters for internal combustion engines                                                        \\
				Engines and Parts             & 841330              & Fuel, lubricating and cooling pumps for motor engines                                                     \\
				Miscellaneous Parts           & 940190              & Parts of seats                                                                                            \\
				Miscellaneous Parts           & 870891              & Radiators for motor vehicles                                                                              \\
				Miscellaneous Parts           & 870892              & Mufflers and exhaust pipes for motor vehicles                                                             \\
				Miscellaneous Parts           & 870893              & Clutches and parts thereof for motor vehicles                                                             \\
				Miscellaneous Parts           & 870894              & Steering wheels, columns \& boxes for motor vehicles                                                      \\
				Miscellaneous Parts           & 870895              & Safety airbags with inflater system; parts thereof   for the motor vehicles of 87.01-87.05                \\
				Miscellaneous Parts           & 870899              & Motor vehicle parts nes                                                                                   \\
				Miscellaneous Parts           & 871690              & Trailer/non-mechanically propelled vehicle parts nes                                                      \\
				Miscellaneous Parts           & 940120              & Seats, motor vehicles               
			\end{longtblr}
			\label{hs07}
			
			\begin{longtblr}[
				caption = {Further products from HS 2012},
				label = {up-down},
				]{
					colspec = {|p{5.5cm}|p{1.5cm}|p{9cm}|},
					rowhead = 1,
					hlines,
					row{even} = {gray9},
					row{1} = {CadetBlue!30},
				} 
				
				\hline
				\textbf{Category}   & \textbf{HS12 code} & \textbf{HS12 description} \\
				\hline
				Electrical and Electric Parts	& 850710	& Electric accumulators; lead-acid, of a kind used for starting piston engines, including separators, whether or not rectangular (including square) \\
				Electrical and Electric Parts	&850750	& Electric accumulators; nickel-metal hydride, including separators, whether or not rectangular (including square) \\
				Electrical and Electric Parts	& 850760	& Electric accumulators; lithium-ion, including separators, whether or not rectangular (including square) \\
				Electrical and Electric Parts	& 850780	& Electric accumulators; other than lead-acid, nickel-cadmium, nickel-iron, nickel-metal hydride and lithium-ion, including separators, whether or not rectangular (including square) \\
			\end{longtblr}
			\label{hs12}
			
			\begin{longtblr}[
				caption = {Further products from HS 2017},
				label = {up-down},
				]{
					colspec = {|p{5.5cm}|p{1.5cm}|p{9cm}|},
					rowhead = 1,
					hlines,
					row{even} = {gray9},
					row{1} = {CadetBlue!30},
				} 
				
				\hline
				\textbf{Category}   & \textbf{HS17 code} & \textbf{HS17 description} \\
				\hline
				Electrical and Electric Parts	& 850710	&Electric accumulators; lead-acid, of a kind used for starting piston engines, including separators, whether or not rectangular (including square)\\
				Electrical and Electric Parts &	850750	&Electric accumulators; nickel-metal hydride, including separators, whether or not rectangular (including square)\\
				Electrical and Electric Parts&	850760	&Electric accumulators; lithium-ion, including separators, whether or not rectangular (including square)\\
				Electrical and Electric Parts	& 850780	& Electric accumulators; other than lead-acid, nickel-cadmium, nickel-iron, nickel-metal hydride and lithium-ion, including separators, whether or not rectangular (including square)\\
				Hybrid Vehicles	& 870340	& Vehicles; with both spark-ignition internal combustion reciprocating piston engine and electric motor for propulsion, incapable of being charged by plugging to external source of electric power\\
				Hybrid Vehicles &	870360	& Vehicles; with both spark-ignition internal combustion reciprocating piston engine and electric motor for propulsion, capable of being charged by plugging to external source of electric power\\
				Hybrid Vehicles &	870350	& Vehicles; with both compression-ignition internal combustion piston engine (diesel or semi-diesel) and electric motor for propulsion, incapable of being charged by plugging to external source of electric power\\
				Hybrid Vehicles &	870370 &	Vehicles; with both compression-ignition internal combustion piston engine (diesel or semi-diesel) and electric motor for propulsion, capable of being charged by plugging to external source of electric power\\
				Electric Vehicles &	870380	& Vehicles; with only electric motor for propulsion\\
			\end{longtblr}
			\label{hs17}


		\end{appendices}
	\end{spacing}
\end{document}